\documentclass[fleqn,usenatbib]{mnras}

\usepackage{newtxtext,newtxmath}

\usepackage[T1]{fontenc}
\usepackage{ae,aecompl}

\usepackage{verbatim}
\usepackage{graphicx}	
\usepackage{amsmath}	
\usepackage{amssymb}	
\usepackage{enumitem}
\usepackage{array}
\usepackage{multirow}
\usepackage{lipsum}
\usepackage{siunitx} 
\DeclareSIUnit\solarmass{M\ensuremath{_\odot}}

\newcommand{\sunmass}{\SI{}{\solarmass}}

\def\lc#1{{\textcolor{black}{\bf #1}}}

\title[Neutron stars mergers in a stochastic chemical evolution model]{Neutron stars mergers in a stochastic chemical evolution model: impact of time delay distributions}

\author[L. Cavallo et al.]{
L. Cavallo,$^{1}$
G. Cescutti$^{2,3}$
and F. Matteucci$^{1,2,4}$
\\
$^{1}$Dipartimento di Fisica, Sezione di Astronomia, Università di Trieste, Via G. B. Tiepolo 11, 34143 Trieste, Italy\\
$^{2}$ INAF, Osservatorio Astronomico di Trieste, via G. B. Tiepolo 11, 34143 Trieste, Italy\\
$^3$ IFPU, Istitute for the Fundamental Physics of the Universe, Via Beirut,  2, 34151, Grignano, Trieste, Italy\\
$^{4}$ INFN, Sezione di Trieste, via A. Valerio 2, 34127 Trieste, Italy
}

\date{Accepted XXX. Received YYY; in original form ZZZ}

\pubyear{2020}

\begin{document}
\label{firstpage}
\pagerange{\pageref{firstpage}--\pageref{lastpage}}
\maketitle

\begin{abstract}
We study the evolution of the [Eu/Fe] ratio in the Galactic halo by means of a stochastic chemical evolution model considering merging neutron stars as polluters of europium.  
We improved our previous stochastic chemical evolution model by adding a time delay distribution for the coalescence of the neutron stars, instead of constant delays.
 The stochastic chemical evolution model can reproduce the trend and the observed spread in the [Eu/Fe] data with neutron star mergers as unique producers if we assume:
 i) a delay time distribution $\propto t^{-1.5}$, 
 ii) a $M_{Eu}= \num{1.5e-6} M_{\odot}$ per event, 
 iii) progenitors of neutron stars  in the range $9-50M_{\odot}$ and
 iv)  a constant fraction of massive stars in the initial mass function (0.02) that produce neutron star mergers. 
 Our best model is obtained by relaxing point iv) and assuming a fraction that varies with metallicity. 
 We confirm that the mixed scenario with both merging neutron stars and supernovae as europium producers can provide a good agreement with the data relaxing the constraints on the distribution time delays for the coalescence of neutron stars. 
 Adopting our best model, we also reproduce the dispersion of [Eu/Fe] at a given metallicity, which depends on the fraction of massive stars that produce neutron star mergers. Future high-resolution spectroscopic surveys, such as 4MOST and WEAVE, will produce the necessary statistics to constrain at best this parameter.
\end{abstract}




\section{Introduction} \label{intro}
The majority of all nuclei that are heavier than the iron-peak element (A $\geq$ 70) are produced by neutron-capture reactions. The neutron capture processes are divided into two different classes: rapid or r-process (neutron capture timescale shorter than $\beta$ decay) and slow or s-process (in this case the neutron-capture timescale is longer than $\beta$ decay). Most neutron-capture elements are produced by both r and s-process, but for some of these heavy nuclei, the production is dominated by only one process. 
A series of works found a spread of r-process elements in the metal-poor environment of the Galactic halo \citep{mcwilliam98,koch2002europium, honda2004spectroscopic, fulbright2000abundances}. This spread can reach 2 dex at [Fe/H]$\sim-3$ dex. On the other hand,  [$\alpha$/Fe] ratios (where $\alpha$ stands for $\alpha$-elements) show a smaller scatter than r-process elements. The $\alpha$-element spread, if real and not due to observational uncertainties, can be due to cosmic selection effects favoring contributions from supernovae in a certain mass range \citep[see][]{ishimaru2003detection,refId0}.
In literature Eu is often indicated as a good r-process tracer for two basic reasons: i) more than 90$\%$ of Eu in the solar system has been produced by r-process \citep{cameron1982elemental, howard1986parametric, 2015MNRAS.449..506B}. ii) Europium is one of the few r-process elements that shows clean atomic lines in the visible part of the electromagnetic spectrum, and this makes Eu abundances easier to measure than other r-process elements \citep{woolf1995r}.\\
Two main astrophysical sites have been proposed for Eu production: i) core-collapse SNe (Type II SNe during explosive nucleosynthesis \citep{cowan1991r, woosley1994r, wanajo2001r}. However, there are still many uncertainties in the physical mechanism involved in Eu production in Type II SNe \citep{arcones2007nucleosynthesis}. ii) neutron star mergers (NSM) can provide a strong Eu production 
\citep{freiburghaus1999r,Wanajo_2014,2008AstL...34..189P,1982ApL....22..143S, 2007A&A...467..395O, 2013ApJ...773...78B, 2013PhRvD..88d4026H, 2014MNRAS.443.3134P}. Each event can produce a total amount of Eu from $10^{-7}$ to $10^{-5} \SI{}{\solarmass}$ \citep{korobkin2012astrophysical}.\\
Previous models, such as \citet{argast2004neutron}, computed the evolution of Eu for the halo of our Galaxy with an in-homogeneous chemical evolution model. They concluded that NSMs cannot be the major production site of Eu due to their low merging rate. In this scenario NSMs failed to reproduce the observation of stars at low metallicity ([Fe/H]< $-2.5$). Later \citet{cescutti2006chemical} found that, in a model with instantaneous mixing, SNe II can be entirely responsible for the production of Eu. Moreover, he suggested that Eu originates from stars in a mass range 12-30 $\SI{}{\solarmass}$. \\
\citet{matteucci2014europium} showed that, in a chemical model with instantaneous mixing approximation (I.M.A), neutron stars (NS) can be the only production site of Eu under some conditions: the time scale of coalescence cannot be longer than 1 Myr; the yield of Eu per single event is around $\num{3e-6}$ $\SI{}{\solarmass}$; the mass range of neutron stars progenitors is 9-50 $\SI{}{\solarmass}$. With similar assumptions on NSM parameters, \citet{cescutti2015role} proved that with a stochastic chemical evolution model these events can also explain the large spread of [Eu/Fe] vs [Fe/H] observed in the halo of our Galaxy. It was also found out that the scenario which best 
reproduces the observational data is the one where both neutron star mergers and a fraction of Type II supernovae produce Eu. A main assumption of the previous models is the short coalescence time of NS systems, but some observational bounds cannot be satisfied by a constant and short coalescence time, such as to explain the recently observed event GW170817 which occurred in an early-type galaxy with no star formation, as well as to reproduce the cosmic rate of short Gamma-Ray Bursts (short-GRBs). Recently \citet{cote2019neutron} proved that, if we assume NSM as the only r-process site, there are some tensions between models and observational data when we drop the condition of short and constant coalescence time. In particular, they found that NSMs with a coalescence time that follows the same delay time distribution (DTD) of SNe Ia cannot reproduce the decreasing trend of [Eu/Fe] at [Fe/H]> $-1$ dex in the Galactic disk. However, \citet{schonrich2019chemical} showed that, also with a DTD for NSM (with a characteristic merger time-scale $t_{NS}$ = 150 Myr), they were able to explain the observed abundance patterns assuming a 2-phase ISM (hot and cold).
On the other hand, \citet{simonetti2019new} adopted  a DTD for NSM built from theoretical considerations and concluded that either SNeII or a fraction of NSM variable in time can potentially explain the [Eu/Fe] in the Galaxy as well as the cosmic rate of short-GRBs.\\
Moreover, the effect of a DTD for NSM on the chemical evolution of r-process elements was also explored by \citet{Shen_2015}. In particular, they investigated the chemical evolution of the heavy r-process elements in our Galaxy using a high-resolution cosmological simulation (Eris) for the formation of a Milky-Way like galaxy. They used a power-law slope with two different exponents: $\propto t^{-x}$ ($x=1; 2$).
Later, in the framework of the hierarchical galaxy formation, \citet{2016ApJ...830...76K} explored the effects of propagation of NSM ejecta across proto-galaxies on the r-process chemical evolution. Considering these effects, they found that NSM with a DTD are able to reproduce the emergence of r-process elements at very low metallicity ([Fe/H] $\sim-$3 dex).\\
In this paper, we want to test whether NSM with coalescence time that follows a proper DTD can explain the spread of [Eu/Fe] of metal-poor stars ([Fe/H] < $-$1 ) in the Galactic halo. To compute the chemical enrichment we adopt a stochastic chemical evolution model, proposed in \citet{cescutti2008}, that mimics an in-homogeneous mixing thanks to a stochastic modeling. We also explore cases in which both NSM (with a DTD) and Type II SNe produce Europium. In particular, in the last part of the work, we take into account the contribution of magneto-rotationally driven (MRD) supernovae in the Eu enrichment. MRD SNe \citep{winteler2012magnetorotationally,2015ApJ...810..109N, 2015Natur.528..376M} have been indicated as a promising source of r-process in the early Galaxy \citep*{cescutti2014explaining}.\\
The paper is organised as follows: in Section \ref{observational_evidence} we describe the observations; in Section \ref{CEM} we introduce the adopted chemical evolution model. In Section \ref{results} we discuss our results and finally in Section \ref{Conclusions} we draw some conclusions. 

\section{OBSERVATIONAL CONSTRAINTS} \label{observational_evidence}
To test the predictions of our model we use the abundances of the halo stars contained in \citet{roederer2014search}. The sample contains abundances of 115 metal-poor stars. We chose to test our models with a data-set provided by a single author even if the dimension of the sample is quite small compared to the total amount of data that are available in literature ($\simeq 400$), e.g. JINAbase \citep{Abohalima_2018}. We have opted for this choice in order to remove potential off-sets between different data.

\subsection{[Eu/Fe] of metal-poor stars in the Galactic halo} \label{[Eu/Fe] halo}
The abundances measured in halo stars show a clear large scatter in the ratio of [r/Fe], where r stands for an r-process element, versus metallicity. \citet{cescutti2008} suggested that the wider spread observed in neutron-capture elements, compared to [$\alpha$/Fe] ratios, is a consequence of the difference in mass ranges between the production sites. This also implies that, in the early Universe, the production of Eu must have been rare and prolific compared to the one of $\alpha$-elements.\\

\subsection{Neutron Star Mergers as progenitors of Short Gamma-Ray Bursts}  \label{SGRB}
Gamma-ray bursts display a bi-modal duration distribution with a separation between the short and long-duration bursts at about 2 s. The progenitors of Long GRBs have been identified as massive stars. On the other hand, Short-GRBs are thought to be correlated with compact object mergers \citep{berger2014short, 1989Natur.340..126E, 2013Natur.500..547T}. This hypothesis has been recently reinforced by the observation of a short GRB, that followed the NSM event GW170817 detected by LIGO/Virgo Collaboration \citep{abbott2017gw170817}. In particular, NGC 4993, the host galaxy of GW170817, is an early-type galaxy \citep{abbott2017search, coulter2017swope}. If we assume a coalescence time constant and short, at least < 10 Myr as suggested in \citet{matteucci2014europium} and \citet{cescutti2015role}, it will be impossible to detect a NSM in an early-type galaxy, where the star formation is over and all the NS-NS systems should have already merged. This requires the adoption of a DTD  including long timescales. As a caveat, we should also point out that it is not impossible that the merger took place in a dwarf galaxy still star forming that we are unable to distinguish.

\begin{figure*}
    \centering
    \includegraphics[trim={2cm 2cm 2cm 2cm}, clip, scale=0.7]{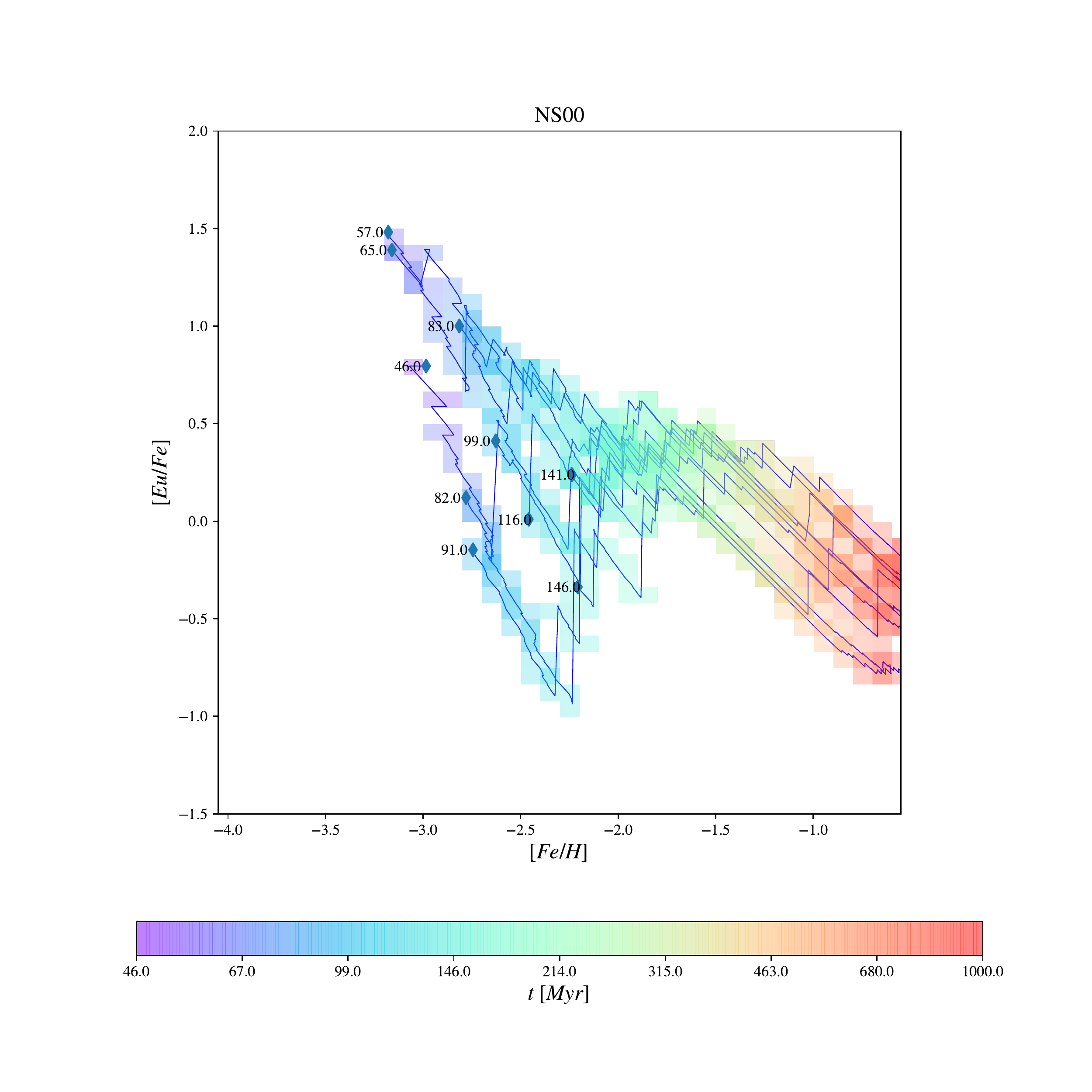}
    \caption{Results of [Eu/Fe] vs [Fe/H] for ten realisations of NS00 model. As reported in Table \ref{tab:Models}, this model has a constant delay time for NSMs of 1 Myr. With the colour map we show the time at which the realisation pass through a certain point in the [Eu/Fe] vs [Fe/H] plane. We also report the initial point and the time at which the first NSM has exploded.}
    \label{fig:graficoNuovo}
\end{figure*}

\section{THE CHEMICAL EVOLUTION MODEL} \label{CEM}
The chemical evolution model adopted here is the same as in \citet{cescutti2015role}, which is based on the stochastic model developed by \citet{cescutti2008}. We review its main characteristics to improve the reader comprehension of the work.\\ 
The Galactic halo is simulated by means of 200 stochastic realisations. Each realisation consists of a non-interacting region with the same typical volume.  The dimensions of the typical volume were chosen in order to neglect the interactions between different regions. In fact, for typical ISM densities, a supernova remnant becomes indistinguishable from the ISM  before reaching $\sim$50 pc \citep{thornton1998energy}. On the other hand, we do not want a too large volume because in that case, we would lose the stochasticity. For these reasons has been chosen a typical volume with a radius of $\sim$90 pc. We consider 200 realisations to ensure a good statistical sample.\\
The model uses time-steps of 1 Myr, which is shorter than any stellar lifetime considered in this model; the minimum lifetime is, in fact, 3 Myr for an 80 $\sunmass$ star, which is the maximum stellar mass considered.\\
In each region, following the homogeneous model by \citet{chiappini2008new}, we assume the following function for the infall of gas with primordial composition:
\begin{equation}
    \frac{dGas_{in}(t)}{dt} = I_{nfall} e^{-(t-t_0)^2 / 2\sigma^{2}_{0}  }
\end{equation}
where $t_0$ is set to 100 Myr, $\sigma_{0}$ is 50 Myr and $I_{nfall}$ is equal to $\num{1.28e4}$ $\sunmass$ $Myr^{-1}$. We define the Star Formation Rate (SFR) as:
\begin{equation}
    SFR(t) = \nu \left( \frac{\sigma_{gas}(t)}{\sigma_h}  \right) ^{1.5}
\end{equation}
where $\sigma_{gas}(t)$ is the surface density of the gas inside a volume at each time-step, $\sigma_h= 80$ $\sunmass$ $pc^{-2}$ and $\nu$ is set to $2862$ $\sunmass$ $Myr^{-1}$ . We also take into account an outflow that follows the law:
\begin{equation}
    \frac{dGas_{out}(t)}{dt} = Wind * SFR(t)
\end{equation}
where $Wind$ is set to 8.\\
In all the subhaloes of the model, we assume the same SFR and infall laws. The following lines will introduce the stochastic part contained in our model.\\
Let us assume that we know the mass that is transformed at each time-step into stars ($M^{new}_{stars}$), then we generate one star with a mass sorted out with a random function, weighted on the initial mass function (IMF) of \citet{scalo1986stellar} in the mass range from 0.1 to 100 $\sunmass$. After that, the mass of the second star is extracted, and so on until the total mass of newborn stars reaches $M^{new}_{stars}$. In this way, the total amount of mass transformed into new stars is the same in each region at each time-step, but the total number and mass distribution of stars is different. For all the stars we also know mass and lifetime. In particular, we assume the stellar lifetime of \citet*{maeder1989grids}.\\
When a star dies, it enriches the ISM with its newly produced elements and with the unprocessed elements present in the star since its birth. Our model considers a detail pollution from SNe core-collapse (M $> 8\sunmass$), AGB stars, NSMs, and SNe Ia, we follow the prescriptions for the single degenerate scenario of \citet{1986A&A...154..279M}.\\\\
\noindent
In Fig. \ref{fig:graficoNuovo}, we present graphically how the chemical enrichment proceeds in our stochastic model . For clarity, only ten realisations are shown on the [Eu/Fe] vs [Fe/H] plane. The model has a constant delay time for NSM of 1 Myr and it is one of model studied in \citet{cescutti2015role}, namely NS00.  
In this Figure, we can appreciate how the time at which the first NSM  explodes and pollute stars with europium, varies among the different realisations. Although the delay between the formation of NS binary and NSM is the same and very short, the formation of a NS binary is stochastic. So, we can have realisations where the first stars present europium after only $\sim 50$ Myr, but also realisations where this happens later at around $\sim 150$ Myr. A short formation delay implies less chemical enrichment of the volume. Therefore, the model results typically lie at high [Eu/Fe] and low [Fe/H], the contrary for longer delays (lower [Eu/Fe] and higher [Fe/H]).\\
In Fig. \ref{fig:graficoNuovo}, the reader can also appreciate the different paths followed by each single realisation in the [Eu/Fe] vs [Fe/H] plane. These paths show some patterns, which can be understood in terms of the enrichment that takes place in that region. For example, when a  realisation moves horizontally towards lower metallicities, there is no events enriching the ISM of iron or europium and the gas is diluted by the infalling gas with primordial composition. Then, when an event produces Fe, the realisation moves to higher metallicities and lower [Eu/Fe] ratios. If a NSMs explodes, the realisation makes a jump towards higher [Eu/Fe] values. In general, the height of these "jumps" varies for different realisations, due to the variable amount of Eu that a single NSM can produce (see Equation \ref{eqn:Random}).

\subsection{Stellar yields for Eu} \label{modelEuYields}
For the Eu production sites, we take into account both NSM and core-collapse SNe. We define three parameters to include the Eu production from NSM \citep{matteucci2014europium}:
\begin{enumerate}[leftmargin=*]
    \setlength\itemsep{7pt}
    \item the fraction of massive stars that generate a binary system of neutron stars that will eventually merge, $\alpha_{NS}$.
    \item the amount of Eu produced by a single merging event, $M^{Eu}_{NS}$.
    \item the delay time between the formation of the binary system and the merging event. From now on we will call it coalescence time, $t_c$.
\end{enumerate}
In our work, we assume that a fixed fraction of massive stars, generated during the simulation, is the progenitor of NSMs. The progenitors are chosen randomly among all the generated massive stars in the mass range 8-50 $\sunmass$. We take the progenitor mass range as suggested in \citet{matteucci2014europium}. We assume a similar $\alpha_{NS}$ to the one contained in \citet{matteucci2014europium} ($\sim$ 0.018), which is in agreement with the present-day neutron star merging rate of our Galaxy calculated by \citet{kalogera2004cosmic} ($\sim$80 Myr$^{-1}$).\\
For the nucleosynthesis of Eu, we use empirical values that have been chosen in order to reproduce the surface abundances of Eu in low-metallicity stars as well as the solar abundances of Eu \citep[see][]{cescutti2006chemical}. These values are consistent with the limits calculated by \citet{korobkin2012astrophysical}, who suggested that a single NSM can produce from $10^{-7}$ to $10^{-5}$ $\sunmass$ of Eu. \\
During the work, we have also considered a non-constant Eu production for a single NSM. In general the variation is unknown, so we assume a range from 1$\%$ of the average Eu ($M^{Eu}_0$) to 200$\%$ of it. Since the total mass of Eu produced should be preserved, the $n^{th}$ star ejects a mass of Eu that follows this equation:
\begin{equation} \label{eqn:Random}
    M^{Eu}_{NS}(n)=M^{Eu}_{0}(0.01+1.98 \cdot Rand(n))
\end{equation}
where $Rand(n)$ is a uniform random distribution in the rage $[0,1]$ \citep[same as in][]{cescutti2014explaining}.\\
For the production of Eu from SNe II we adopt yields 
similar to those of \citet{matteucci2014europium} (Mod2SNNS Model). Since recent results showed that the conditions during a supernova Type II explosion may not be able to produce much Eu \citep{arcones2007nucleosynthesis, 2011JPhCS.312d2008W}, we also tested an alternative channel: the magneto-rotationally driven (MRD) SNe. MRD SNe are a particular class of core-collapse supernovae. Here, we assume that the 10$\%$ of the CC-SNe explode as MRD.  This r-process site is active only at low metallicity (Z<$10^{-3}$), so it affects the model results only at low metallicity. These assumptions are identical to the ones contained in \citet{cescutti2015role}. This particular fate is rare, only few SNe explode as MRD-SNe, and as mentioned in \citet{winteler2012magnetorotationally}, it should be more likely to happen at low metallicity \citep{2006A&A...460..199Y}.

\begin{table}
    \centering
    \renewcommand{\arraystretch}{1.2}
\begin{tabular}{c|c c c}

\multicolumn{1}{c|}{\multirow{2}{*}{DTD}} & \multicolumn{3}{c}{percent of NSMs exploded before} \\ 
\multicolumn{1}{c|}{}                     & \multicolumn{1}{c}{$10$ Myr} & \multicolumn{1}{c}{$100$ Myr} & 
\multicolumn{1}{c}{$1000$ Myr} \\ \hline
                   &     &     &\\[-1em]
$\propto t^{-1}$  & 25 $\%$  &  50 $\%$  &   75 $\%$  \\
$\propto t^{-1.5}$ & 69 $\%$  &  91 $\%$  &   98 $\%$ 
\end{tabular}
    \caption{Percentage of NSM already merged at different times for DTD of different shapes. Those values are for DTDs with a $t_c^{min}=1$ Myr}
    \label{tab:vsDTD}
\end{table}

\begin{table*}
\centering
\begin{tabular}{c| c c c c c c}
  & & & & & \\
  Model Name &  DTD & $t^{min}_{c}$ [Myr] & $\alpha_{NS}$ & $M^{Eu;NSM}_0$ [$\sunmass$] & $M^{Eu;MRD}_0$ [$\sunmass$] \textsuperscript{$\Psi$} \\ 
  & & & & & & \\
  \hline
  \hline
  & & & & & & \\
  NS00$^a$ & no & 1 & 0.02 & $\num{1.5e-6}$ (varying as eq. \ref{eqn:Random})& no production\\
  NS01$^a$ & " & 10 & " & " & "\\
  NS02$^a$ & " & 100 & " & " & "\\
   & & & & & \\
   \hline
  & & & & & \\
  NSt1$^b$ & $\propto t^{-1}$ & 1 & 0.02 & $\num{4e-6}$ (varying as eq. \ref{eqn:Random}) & no production \\
  NSt2$^b$ & " & 10 & " & " & " \\
  NSt3$^b$ & $\propto t^{-1.5}$ &  1  & " & $\num{1.5e-6}$ (varying as eq. \ref{eqn:Random}) & " \\
  NSt4$^b$ & " & 10 &  " & " & " \\
   & & & & & \\
   \hline
  & & & & & \\
  NS+MRD00$^c$ & no & 1 & 0.02 & $\num{0.6e-6}$ (varying as eq. \ref{eqn:Random}) & $\num{0.6e-6}$ (varying as eq. \ref{eqn:Random})\\
  NS+MRD01$^c$ & " & 10 & " & " & " \\
  NS+MRD02$^c$ & " & 100 & " & " & " \\
 & & & & & \\
   \hline
  & & & & & \\
  NS+MRDt1$^d$ & $\propto t^{-1}$ & 1 & 0.02 & $\num{0.6e-6}$ (varying as eq. \ref{eqn:Random}) & $\num{0.6e-6}$ (varying as eq. \ref{eqn:Random})\\
  NS+MRDt2$^d$ & " & 10 & " & " & " \\
  NS+MRDt3$^d$ & $\propto t^{-1.5}$ & 1 & " & " & " \\
  NS+MRDt4$^d$ & " & 10 & " & " & " \\
   & & & & & \\
   \hline
  & & & & & \\
  NSt3+$\alpha^e$ & $\propto t^{-1.5}$ & 1 & varying as eq. \ref{alpha} ($\overline{\alpha}_{NS}=0.275$)& $\num{2e-6}$ (varying as eq. \ref{eqn:Random}) & no production\\
  NSt1+$\alpha^e$ & $\propto t^{-1}$ & 1 & " & " & " \\
     & & & & & \\
   \hline
  & & & & & \\
  Test1$^i$ & $\propto t^{-1.5}$ & 1 & varying as eq. \ref{alpha} ($\overline{\alpha}_{NS}=0.275$) & $\num{2e-6}$ (varying as eq. \ref{eqn:Random}) & no production\\
  Test2$^i$ & " & " & varying as eq. \ref{alpha} ($\overline{\alpha}_{NS}=0.315$) & $\num{0.8e-6}$ (varying as eq. \ref{eqn:Random}) & " \\
  Test3$^i$ & " & " & varying as eq. \ref{alpha} ($\overline{\alpha}_{NS}=0.355$) & $\num{0.4e-6}$ (varying as eq. \ref{eqn:Random}) & " \\
  \end{tabular}
  \caption{This table summarize the parameters of the models that we test during this work. It is organised as follows: in column 1, we report the name of the model, in column 2, the assumed DTD for coalescence time, in column 3, the minimum delay time for NSM, in column 4, the assumed fraction of massive star that could lead to NSM, in column 5, the assumed yield for NSM, in column 6, the assumed yield for MRD SNe. $^{\Psi}$ When we take into account the Eu production by MRD-SNe we set $\alpha_{MRD}$=0.10.}
  \label{tab:Models}
\end{table*}

\subsection{The coalescence time distribution for NSM}
In this section, we present the different types of coalescence time scale for NSM that we consider in our models. The DTD functions assumed in this work are $\propto t^{-1}$ and $\propto t^{-1.5}$ and defined as follows:
\begin{equation} \label{DTDs}
\begin{split}
DTD(t) = &
\begin{cases} 
0 & \mbox{if } t<t_{min}^c\\ 
A_x t^{-x} & \mbox{if }t_{min}^c <t<10 \mbox{ Gyr} \\
0 & \mbox{if } t>10 \mbox{ Gyr}\\
\end{cases}\\
&with \, x=\{1,1.5\}\, and \, A_x = 1/ \int  \tau^{-x} d\tau;
\end{split}
\end{equation}
where $t_{min}^c$ is the minimum coalescence time (in our models can assume two values: $1$ and $10$ Myr), and $A_x$ is the normalisation constant.\\
We also discuss possible tensions with observations.

\begin{figure*}
  \includegraphics[trim={2cm 1cm 2cm 4cm}, clip, width=\textwidth]{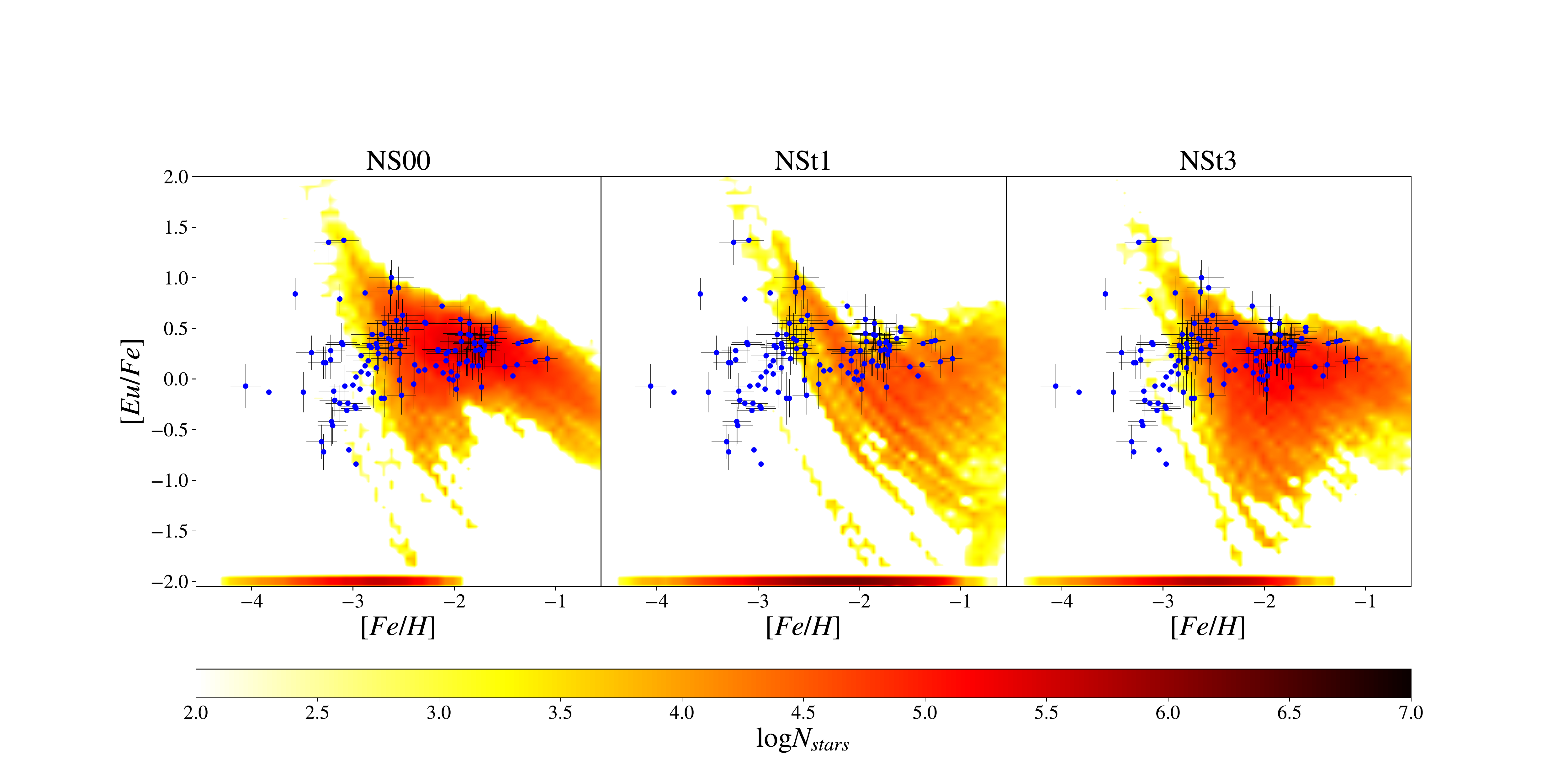}
  \caption{\textit{Left panel}: results of [Eu/Fe] vs [Fe/H] for model NS00. This model has a constant delay time for NSMs of 1 Myr, Eu production that vary as equation \ref{eqn:Random} and a mean value of $\num{1.5e-6}$ \sunmass, no Eu production from CC-SNe. Model NS00 is the same as NS00 contained in \citet{cescutti2015role}. \textit{Central panel}: same as \textit{left panel} but for model NSt1. The only difference from the previous model is the assumption on the coalescence time; in fact in this case we assume a delay time with a DTD $\propto t^{-1} $for NSMs. \textit{Right panel}: same as previous panels but for model NSt3. In this model for the coalescence time of NSMs we assume a DTD $\propto t^{-1.5}$. Note that all the models contained in this figure has minimum coalescence time set to 1 Myr. The long-living stars formed without Eu (formally [Eu/Fe] = $-\infty$) are shown at [Eu/Fe] = $-2.0$ dex.}
  \label{fig:Models1}
  
\end{figure*}

\subsubsection{Neutron star mergers with constant and short coalescence timescale}
As we mentioned in the Introduction, NSM with a short and constant coalescence delay are able to reproduce the decreasing trend of [Eu/Fe] (also called knee) starting from [Fe/H]   $\sim -$1  observed in the Galactic disk \citep{matteucci2014europium} and present also for $\alpha$-elements. As showed by \citet{cescutti2015role}, they can also explain the [Eu/Fe] spread in metal-poor stars in the Milky Way halo. However, a short and constant coalescence time is incompatible with several observations \citep[see ][]{simonetti2019new,cote2019neutron}. To begin with, if we assume that NSMs are progenitors of short-GRBs \citep{berger2014short}, they cannot explain the observation of short GRBs in early-type galaxies, where star formation has stopped several Gyr ago. Furthermore, a short coalescence timescale (<100 Myr) is inconsistent with the theoretical estimation of merging times of the seven known NS-NS binary systems, indeed their coalescence timescale ranging from 86 to 2730 Myr  \citep{tauris2017formation}. 
Finally, as already mentioned, a NSM scenario with short and constant timescales cannot explain the event GW170817 observed in an early-type galaxy.

\subsubsection{Neutron star mergers with a DTD $\propto$ $t^{-1}$ } \label{DTD1}
In Literature, a lot of authors have derived the DTD function of SNe Ia from observations. Most of the studies suggest that, SNe Ia follow a DTD with the form $\propto t^{-1}$ \citep[see][]{totani2008delay,maoz2010supernova, graur2011supernovae,maoz2012type, rodney2014type}. This slope is also in agreement with predictions from population synthesis models. Similar techniques can be applied to derive the DTD of short-GRBs (i.e. the DTD of their progenitors: the NSMs).\\
\citet{Fong_2017} found that, the DTD of short-GRBs can have the form of $t^{-1}$. A power-law with a $-1$ slope is also in agreement with population synthesis studies \citep[see][]{Dominik_2012,chruslinska2018double}.
Assuming a similar DTD for NSMs and SNe Ia (i.e. $\propto t^{-1}$) is also consistent with the fact that SNe Ia and short GRBs are detected in similar proportion in early-type galaxies. However, with this assumption on the DTD of NSMs, it was already shown that NSM cannot reproduce the decreasing trend of [Eu/Fe] in the Galactic disk \citep[see][]{cote2019neutron,simonetti2019new}.\\
In our work, we tested this functional form for the DTD with three different lower bounds in the coalescence time: 1 Myr, 10 Myr, and 100 Myr. In order to include the coalescence timescales of NS-NS systems contained in \citep{tauris2017formation}, we should have chosen an upper limit equal to $\infty$. In this work, we choose an upper limit of 10 Gyr because if we assume a larger one it would have changed only the normalization of the DTD, without a significant impact on the results.

\begin{figure*}
  \includegraphics[trim={2cm 4cm 2cm 4cm}, clip, width=\textwidth]{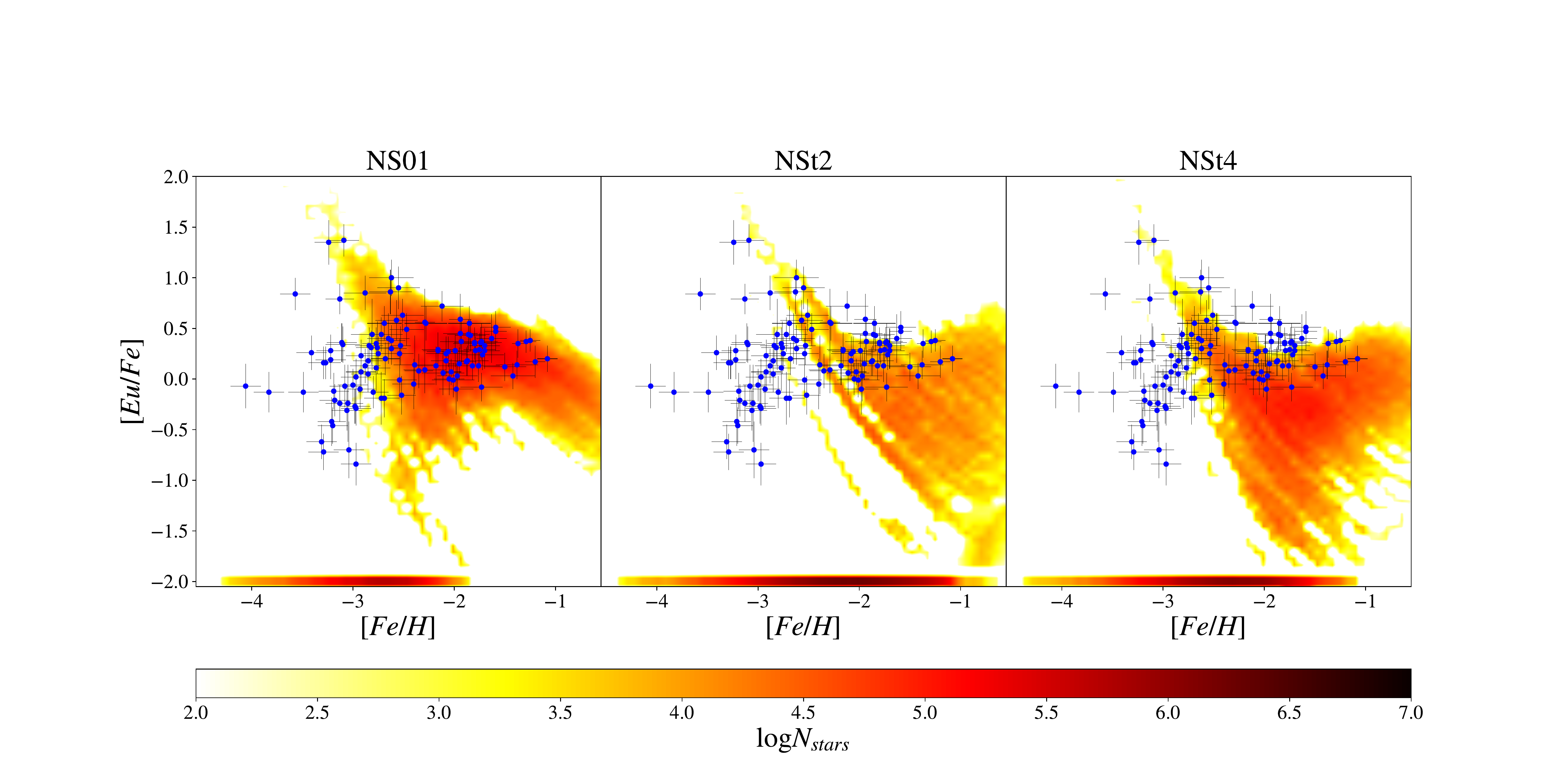}
  \caption{\textit{Left panel}: results of [Eu/Fe] vs [Fe/H] for model NS01. This model has a constant delay time for NSMs of 10 Myr, Eu production that vary as equation \ref{eqn:Random} and a mean value of $\num{1.5e-6}$ \sunmass, no Eu production from CC-SNe.  Model NS01 is the same as NS01 contained in \citet{cescutti2015role}.\textit{Central panel}: same as \textit{left panel} but for model NSt2. The only difference from the previous model is the assumption on the coalescence time; in fact in this case we assume a delay time with a DTD $\propto t^{-1} $for NSMs. \textit{Right panel}: same as previous panels but for model NSt4. In this model for the coalescence time of NSMs we assume a DTD $\propto t^{-1.5}$. Note that all the models contained in this figure has minimum coalescence time set to 10 Myr.}
  \label{fig:Models2}
\end{figure*}

\begin{figure*}
  \includegraphics[trim={2cm 4cm 2cm 4cm}, clip, width=\textwidth]{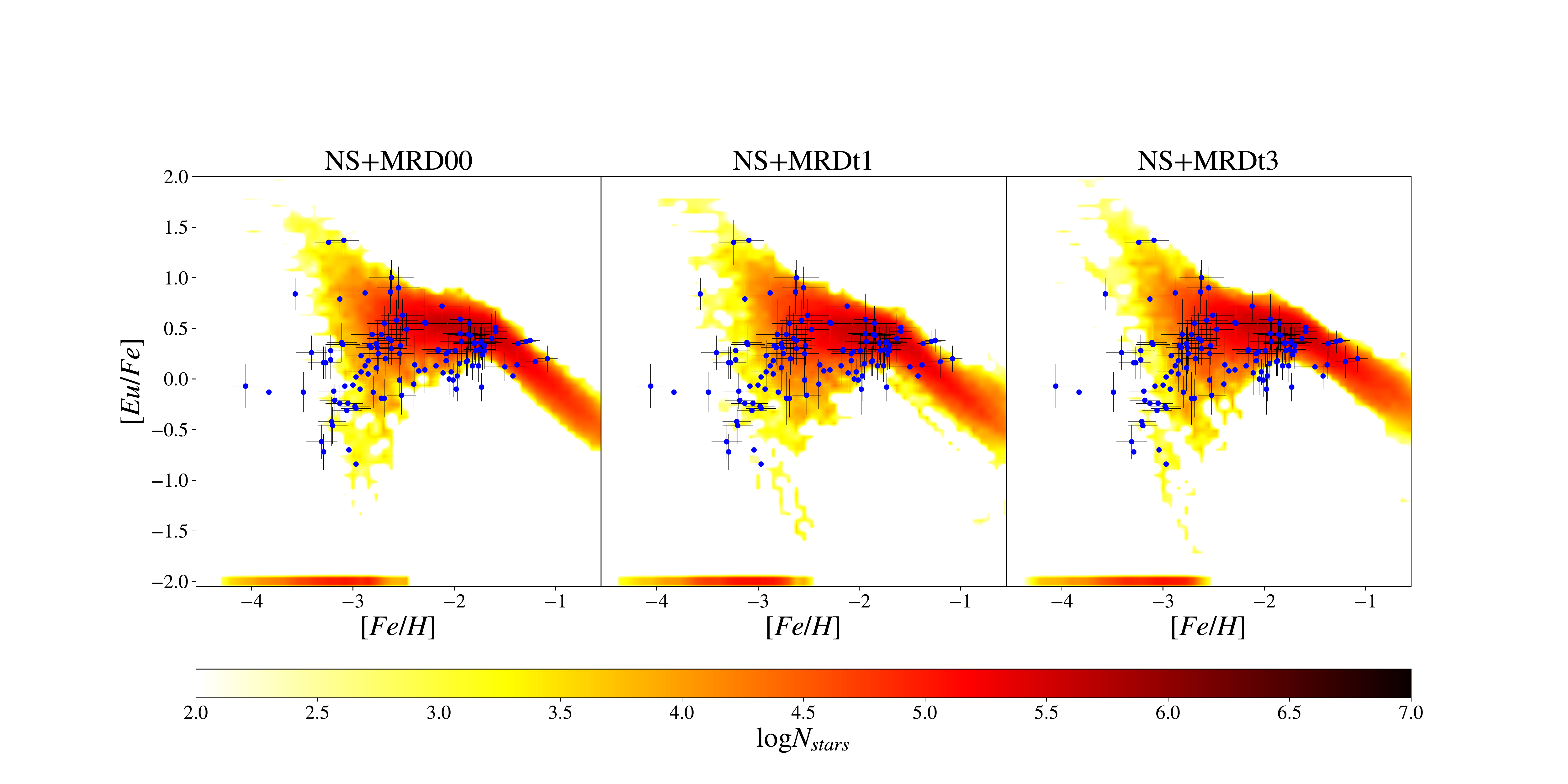}
  \caption{\textit{Left panel}: results of [Eu/Fe] vs [Fe/H] for model NS+MRD00. This model has a constant delay time for NSMs of 1 Myr, Eu production from NSMs that varying as equation \ref{eqn:Random} and a mean value of $\num{0.6e-6}$ \sunmass, Eu production from MRD-SNe (10$\%$ of CC-SNe only at Z<$10^{-3}$) that vary as equation \ref{eqn:Random} and a mean value of $\num{0.6e-6}$ \sunmass. \textit{Central panel}: same as \textit{left panel} but for model NS+MRDt1. The only difference from the previous model is the assumption on the coalescence time; in fact in this case we assume a delay time with a DTD $\propto t^{-1} $ for NSMs with. \textit{Right panel}: same as previous panels but for model NS+MRDt3. In this model for the coalescence time of NSMs we assume a DTD $\propto t^{-1.5}$. Note that all the models contained in this figure has minimum coalescence time set to 1 Myr.}
  \label{fig:Models3}
\end{figure*}

\subsubsection{Neutron star mergers with a DTD $\propto$ $t^{-1.5}$} \label{DTD1.5}
We also tested a DTD $\propto t^{-1.5}$. This kind of slope is consistent with the distribution function of short-GRBs derived by \citet{d2015short}. In particular, a steeper DTD function of the form of $t^{-1.5}$ is not in agreement with the fact that the observed fractions of short-GRBs and SNe Ia are similar. This disagreement could be eased if the DTD function of SNe Ia has also a $t^{-1.5}$ form, as suggested by \citet{heringer2016type}, which showed that SNe Ia follow a DTD with a power-law slope in the range from $-1.3$ to $-1.7$. On the other hand, in the environment of a chemical evolution model, SNe Ia with a DTD $\propto t^{-1.5}$ are not able to reproduce all the [X/Fe] vs [Fe/H] trends in the Galaxy since, with a DTD $\propto t^{-1.5}$, the explosion time-scales of SNe Ia are too short \citep[see][]{10.1111/j.1365-2966.2006.10848.x}.\\
In the light of what we discuss in Sect. \ref{DTD1}, this DTD is not in agreement with the one for SNe Ia and provides coalescence timescales that are variable but still short. In fact, as shown in Table \ref{tab:vsDTD}, more than 90$\%$ of NSMs explode before 100 Myr.

\section{Results} \label{results}
\begin{figure*}
  \includegraphics[trim={2cm 1cm 2cm 2cm}, clip, width=\textwidth]{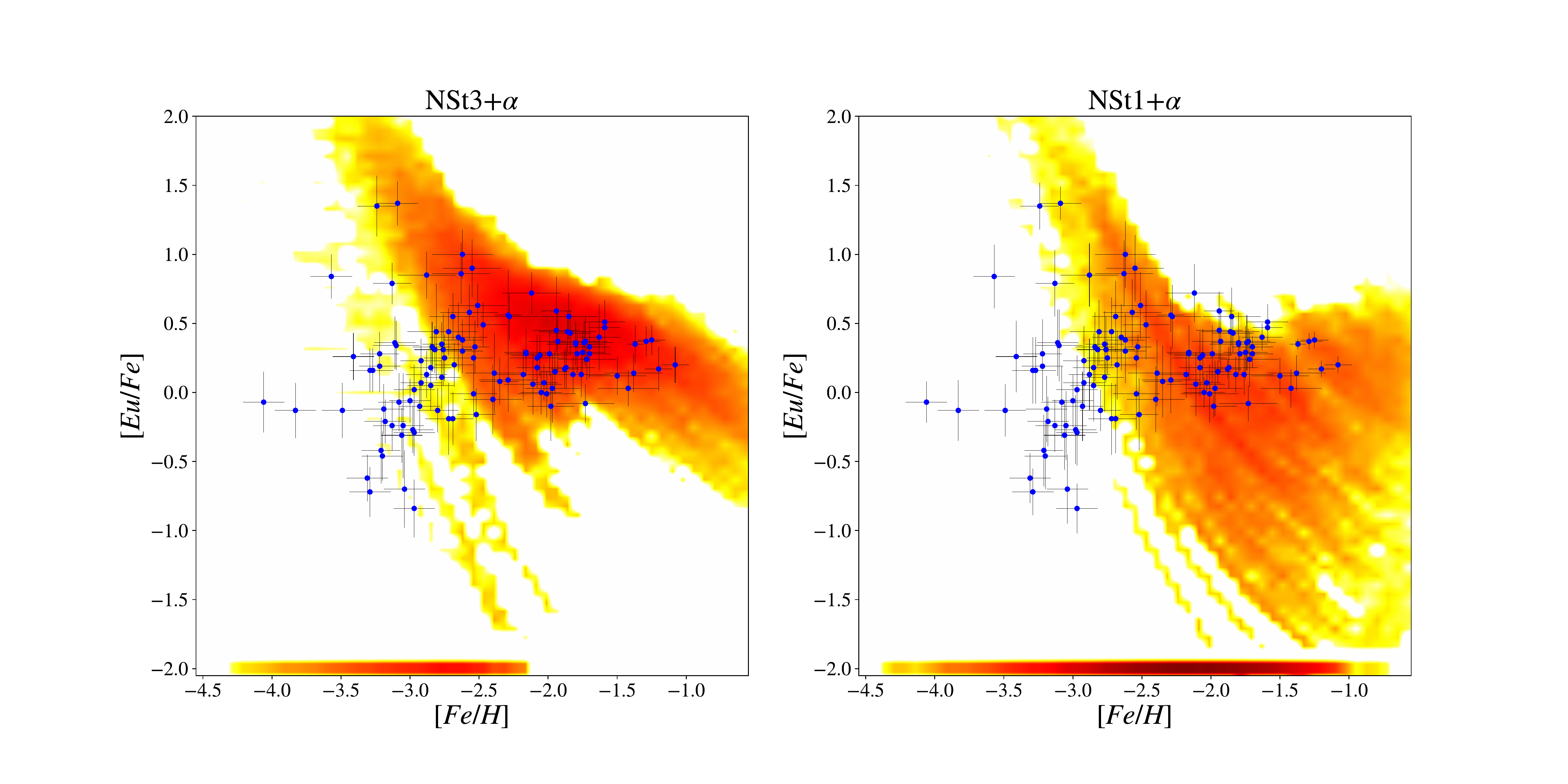}
  \caption{\textit{Left panel}: results of [Eu/Fe] vs [Fe/H] for model NSt3+$\alpha$. Comparing this panel with the right one of Fig. \ref{fig:Models1} is clear that a variable $\alpha_{NS}$ has a great impact on the stars distribution predicted by our models. In particular, an $\alpha_{NS}$ that depends on the metallicity allows models to generate stars Eu-enriched at lower metallicity. \textit{Right panel}: same as \textit{left panel} but for model NSt1+$\alpha$. In this case a variable $\alpha_{NS}$ has the same effect and improves the compatibility between model results and observational data in the metallicity range $-$2.8 < [Fe/H] < $-$2.5.}
  \label{fig:Model4}
\end{figure*}

In the following, we summarise the results of the models we computed, as shown in Table \ref{tab:Models}. They are distinguished in four classes. $a)$ In this scenario only NSM can produce Eu. In this class, is also assumed a constant coalescence time. $b)$ These models test the effects of different DTDs on europium enrichment in the a-class scenarios. $c)$ There both NSM and MRD SNe are r-process sites. For NSM systems we still assume a constant coalescence time. We also assume that, at metallicity (Z<$\num{e-3}$), 10$\%$ of CC-SNe explode as a MRD. $d)$ There we test the effects of relaxing constancy of coalescence time in a NS+MRD scenario. We assume the same $\alpha$=0.1 for MRD. $e)$ With these models we test a variable $\alpha_{NS}$ versus [Fe/H] in a NS-only scenario. $i)$ Last, we test the dependence between the value of $\alpha_{NS}$ and the dispersion of the results at moderate metallicity ($\sim-1.5$ dex). In Figure \ref{fig:Models1}; \ref{fig:Models2}; \ref{fig:Models3}; \ref{fig:Model4}; \ref{fig:Model5} are shown the results, in the [Eu/Fe]vs[Fe/H] plane, from our models. In the plots, at [Eu/Fe] = $-2.0$ dex, we also report the long-living stars formed without Eu (formally [Eu/Fe] $=-\infty$).

\subsection{Models with only NSM} \label{resultNSM}
In Fig. \ref{fig:Models1} is shown the distribution of the long-living stars in the [Eu/Fe]-[Fe/H] plane, as predicted by our stochastic models with the following assumptions: i) Eu is produced only from NSM whose progenitors are in the mass range from 9 to 50 $\sunmass$. ii) The amount of Eu produced from a single event follows equation \ref{eqn:Random} with an average value $(M_0^{Eu})$ of $\num{1.5e-6}$ $\sunmass$. iii) $2\%$ of massive stars are in binary systems with the right characteristics to lead to NSM. iv) The minimum value for the coalescence time is fixed at 1 Myr. The plotted models are NS00, NSt1 and NSt3 (cfr. Table \ref{tab:Models}). In the left panel of Fig. \ref{fig:Models1} is seen that the NS00 model is in agreement with the data for stars with [Fe/H]$>-$3  but it cannot explain the presence of stars with [Eu/Fe]<0 for [Fe/H]$<-$3. Finally, the model does not predict stars with [Eu/Fe]<$-$0.1 at [Fe/H]<$-$3.\\
The reasons of the peculiar diagonal shape in the model results from high [Eu/Fe] with low [Fe/H] to low [Eu/Fe] with higher [Fe/H] (described in \citet{cescutti2015role}) are the following: the upturn in [Eu/Fe], visible at low metallicities, is a consequence of the fixed amount of Eu produced by NSM, coupled with the paucity of NSM events and the constant mixing volume assumed in our model. When a NSM pollutes a simulated box early on, it produces a value in the [Eu/Fe] vs. [Fe/H] space, dependent on the mass of the previous enriching SNeII. The volume enriched by NSM and SNeII with the lowest amount of iron creates the upper tip of this upturn towards low metallicity. Then in all the volumes polluted by NSM, the probability of having another Eu enrichment is low, so they evolve towards lower [Eu/Fe] and higher [Fe/H] by the subsequent enrichment of Fe by SNeII, creating the diagonal shape from high [Eu/Fe] with low [Fe/H] to low [Eu/Fe] with higher [Fe/H]. Indeed, the model struggles to reproduce the stars with [Eu/Fe]< 0 dex at the lowest metallicities. We will examine in details this problem in a future work.\\
When we drop the constancy of the coalescence time and we use a DTD $\propto t^{-1}$ the situation is even worse. In the model NSt1 (middle panel)  fails to reproduce the distribution of the observational data. Finally,
when we take into account a DTD $\propto t^{-1.5}$ (NSt3) we obtain similar results of NS00 model. In fact, as shown in Fig. \ref{fig:Models1} (right panel), the model cannot explain the presence of stars with [Eu/Fe]<$-$0.4 at [Fe/H]<$-$2.8. NSt3 also predicts stars with [Eu/Fe]<-0.2 in the metallicity range $-$2.0<[Fe/H]<$-$1.0. We should notice that results from NS00 and NSt3 are similar. This is due to the fact that a power-law with a $-$1.5 slope keeps the coalescence times short even if they are not constant as mentioned in Sect. \ref{DTD1.5}).\\
The situation does not change if we assume a minimum coalescence time of 10 Myr (Fig. \ref{fig:Models2}).  In Fig. \ref{fig:Models2} we show the results by models with the following assumptions: i) Eu is produced only from NSM and their progenitors are in the mass range from 9 to 50 $\sunmass$. ii) The amount of Eu produced from a single event follows equation \ref{eqn:Random} with an average value $(M_0^{Eu})$ of $\num{1.5e-6}$ $\sunmass$. iii) $2\%$ of massive stars are in binary systems with the right characteristics to lead to merging NS. iv) The minimum value for the coalescence time is fixed at 10 Myr. The plotted models are NS02, NSt2 and NSt4 (see Table \ref{tab:Models}). In the left panel of Fig. \ref{fig:Models2} we can notice that the model NS01 does not predict the presence of stars with [Eu/Fe]<0.3 dex for metallicity lower than $-$2.8 dex. On the other hand, there is good agreement with the observed europium abundances of stars with [Fe/H]>$-$2.7 dex.\\
For model NSt2 (central panel of Fig. \ref{fig:Models2}) the situation is similar to model NSt1 (central panel of Fig. \ref{fig:Models2}). Again, when we drop the constancy of the coalescence time and we also assume a DTD $\propto t^{-1}$, models completely fail to reproduce the observational data. Finally, model NSt4, which assumes a DTD $\propto t^{-1.5}$, fails to reproduce europium abundances of stars with [Fe/H] $<-$2.8 dex. With a mean value of $\num{1.5e-6} \sunmass$ for the Eu production, it also cannot reproduce some stars of the upper envelope of the observed star distribution.

\subsection{Models with NSM and MRD-SNe} \label{resultMRD}
As seen in the previous section, a scenario, where NSMs are the only r-process site, fails to predict the presence of Eu in stars with metallicity [Fe/H] < $-$2.8 dex, even if we assume a constant and short delay time with our stochastic model. A scenario where CC-SNe are the only r-process site is not supported by nucleosynthesis models \citep[see][]{arcones2007nucleosynthesis, Arcones_2012}: in particular, neutrino winds in SNe II explosions are proton-rich and therefore they struggle to produce the heaviest neutron-capture elements (such as Eu). On the other hand, \citet{Siegel_2019} noticed that collapsar (collapse of rotating massive stars) accretion disks also produce neutron-rich outflows that synthesize heavy r-process nuclei, despite the comparatively proton-rich composition of the infalling star. In Sect. \ref{modelEuYields} we discussed an alternative channel for the Eu production: the Magneto-Rotational Driven (MRD) SNe. Now, we want to test if a scenario where both NSMs and MRD-SNe can produce Eu is able to reproduce the abundance of Eu the halo stars. \\
In Fig. \ref{fig:Models3} we report the results of our models with the following assumptions: i) Eu is produced both from NSMs and MRD-SNe. ii) the progenitors of NSMs are in the mass range from 9 to 50 $\sunmass$. ii) The amount of Eu produced from a single NSM event follows equation \ref{eqn:Random} with an average value $M_0^{Eu;NSM} =\num{0.6e-6}$ $\sunmass$. v) $2\%$ of massive stars are in binary systems with the right characteristics to lead to merging NS. iv) At low metallicity (Z<$10^{-3}$), 10 $\%$ of CC-SNe explode as MRD. v) the amount of Eu produced by a single MRD explosion follows equation \ref{eqn:Random} with an average value $M_0^{Eu;MRD}=\num{0.6e-6}$ $\sunmass$ (same as NSMs). The plotted models are NS+MRD00, NS+MRDt1 ,and NS+MRDt3 (cfr. Table 2). \\
Model NS+MRD00 (right panel of Fig. \ref{fig:Models3}) is in good agreement with the observational data and it well predicts the presence of stars with [Fe/H] < $-$3 dex and [Eu/Fe]<0 dex. 
Moreover, also assuming the two different DTDs for NSM, we can still reproduce the data, see Fig. \ref{fig:Models3} middle and right panels.  This is not surprising, since the model at low metallicity, in this case, is basically enriched by MRD SNe. It has been shown also in \citet{cescutti2015role}; in that paper they used a delay for NSM of 100 Myr, $\alpha_{NS}$=0.02, every NSM producing a constant amount of Eu equal to $\num{1.5e-6}$ $\sunmass$ and a single MRD SN produces, on average, $\num{1e-6}$ $\sunmass$ and 10$\%$ of stars in the mass range $8-80$ $\sunmass$ explodes as MRD SNe. 

\begin{figure*}
  \includegraphics[trim={2cm 4cm 2cm 4cm}, clip, width=\textwidth]{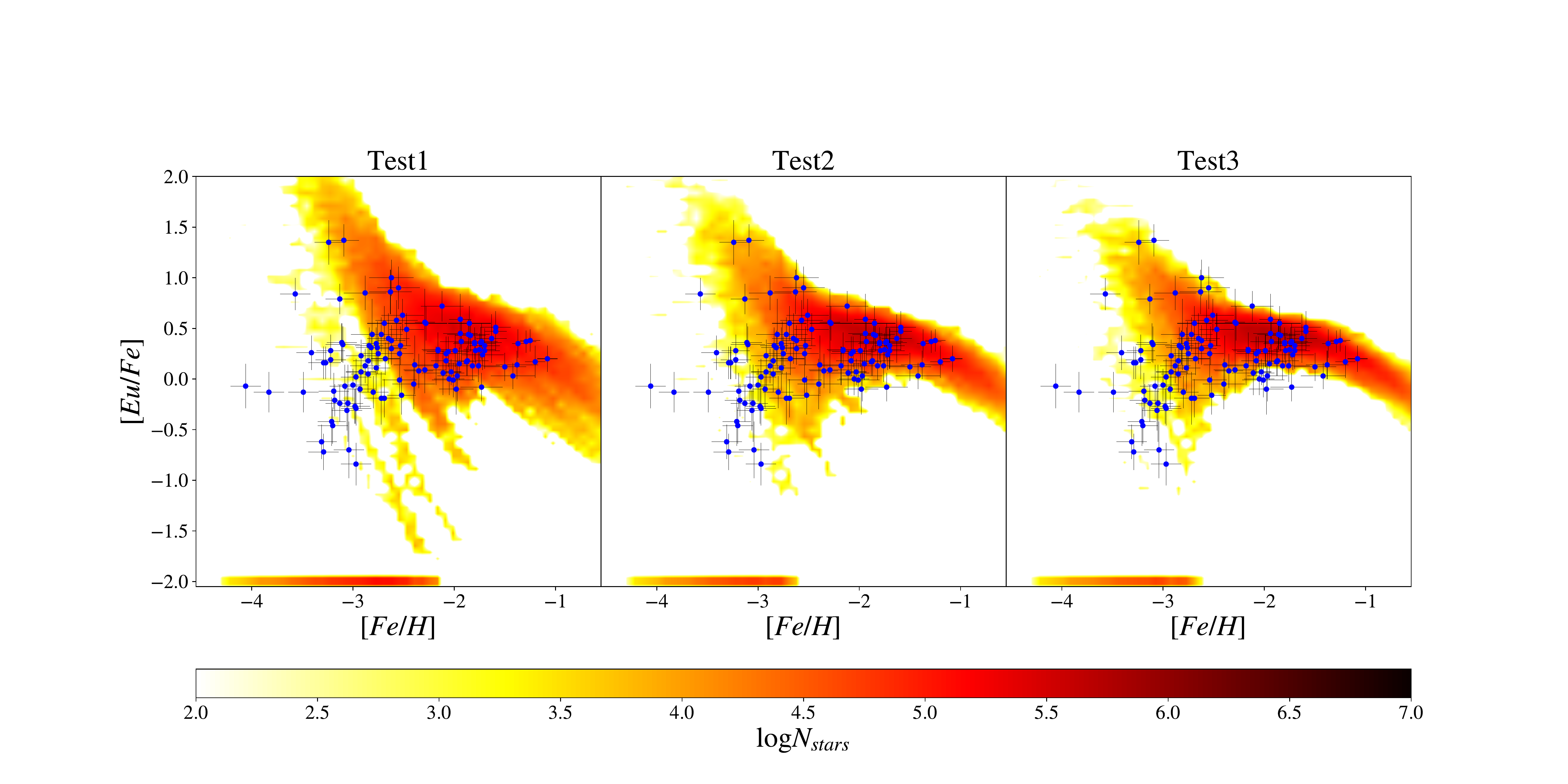}
  \caption{\textit{Left panel}: results of [Eu/Fe] vs [Fe/H] for model Test1. This model is identical to NSt3+$\alpha$ model. We plot it again to emphasise the consequences of the variation of $\alpha_{NS}$ and $M_0^{Eu}$. \textit{Central panel}: same as \textit{left panel} but for model Test2. In this model, the equation \ref{alpha} is up-shifted  by 0.04, the function shows two plateau with $\alpha_{NS}$=0.315 and 0.06. In order to maintain constant the total amount of produced Eu, we reduce $M_0^{Eu}$ to $\num{0.8e-6} \sunmass$. \textit{Right panel}: same as \textit{left panel} but for model Test3. This model, the $\alpha_{NS}$ versus [Fe/H] relation, is up-shifted by 0.08 therefore the function, described by equation \ref{alpha}, shows two plateau with $\alpha_{NS}$=0.355 and 0.1. In this case $M_0^{Eu}$ is reduced to $\num{0.4e-6} \sunmass$. }
  \label{fig:Model5}
\end{figure*}

\begin{figure}
    \centering
    \includegraphics[width=\columnwidth, trim= 0cm 0cm 1cm 0cm, clip]{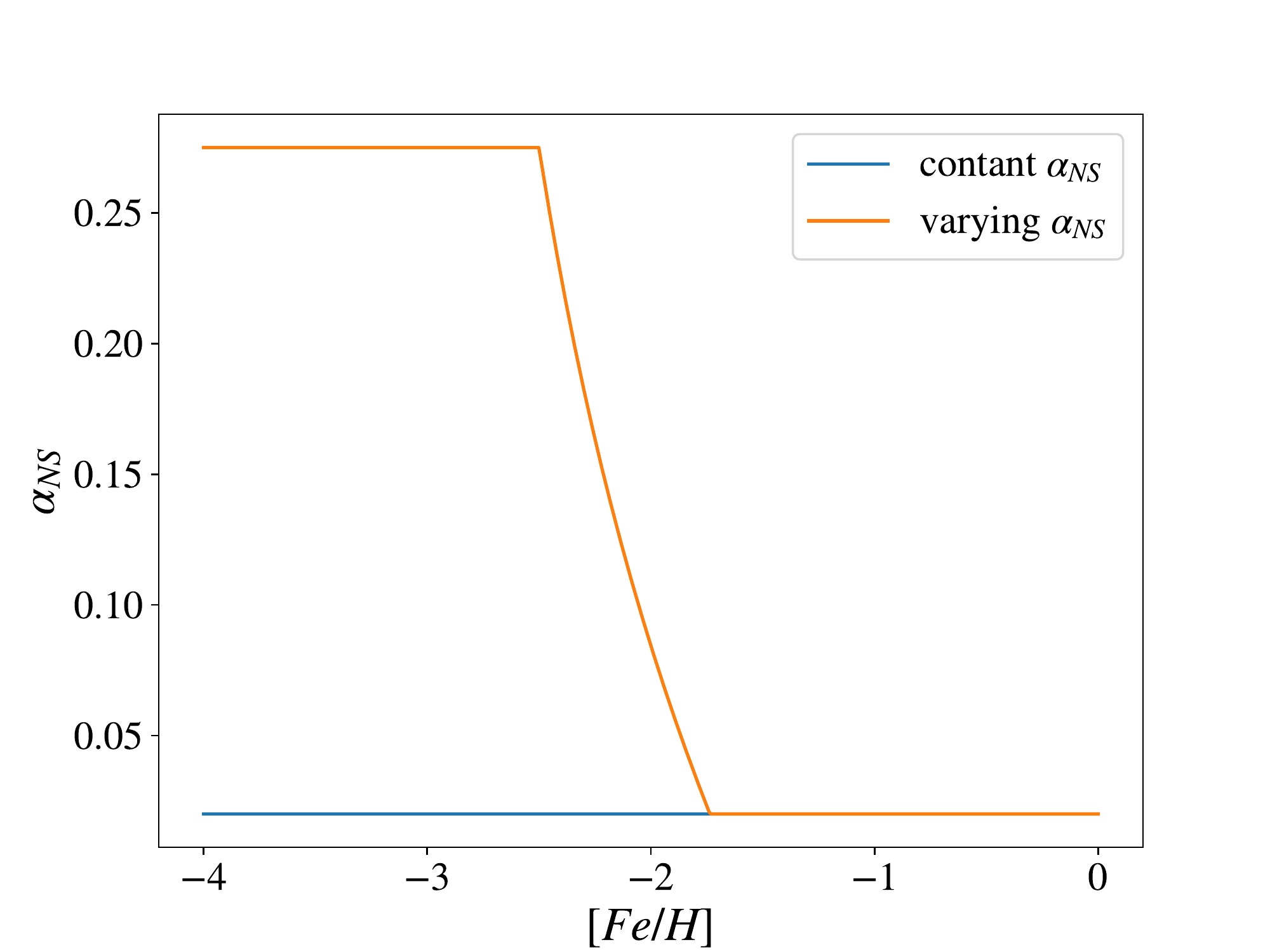}
    \caption{The evolution with metallicity of $\alpha_{NS}$ in two different scenarios.}
    \label{fig:alpha}
\end{figure}

\subsection{Models with variable \texorpdfstring{$\alpha_{NS}$}{TEXT}} \label{Modelli_alpha}
Another possible way to solve it is to relax the assumption of constancy of the fraction of massive stars that can generate a binary system of neutron stars which will eventually merge, $\alpha_{NS}$. \\
Several works investigate the formation of double NS systems \citep{bogomazov2007evolution,ivanova2008formation, mennekens2014massive, shao2018black}. In particular, as shown in \citet{giacobbo2018progenitors}, metallicity plays a crucial role in the formation of binary systems of compact objects. For these reasons, we decide to test this scenario with our chemical evolution model.\\
We assume a dependence of $\alpha_{NS}$ on [Fe/H] (see Fig. \ref{fig:alpha}) similar to the one assumed in model 4AV2 contained in \citet{simonetti2019new}. With this assumption $\alpha_{NS}$ varies as
\begin{equation} \label{alpha}
\alpha_{NS}(\text{[Fe/H]}) = \begin{cases} \overline{\alpha}_{NS} & \mbox{if } \text{[Fe/H]} \leq -2.5 \\ \overline{\alpha}_{NS}(1-ln(\text{[Fe/H]}+z_0)+z_1) & \mbox{if } \text{[Fe/H]} > -2.5 \\
\alpha^{min}_{NS} & \mbox{if } \alpha_{NS}<\alpha^{min}_{NS} \\
z_0=3.0 dex; z_1=ln(0.5) & \\
\end{cases}
\end{equation}
\noindent
In order to test this scenario we have built a new model (NSt3+$\alpha$) with the following assumptions: i) Eu is produced only from NSM, whose progenitors are in the mass range from 9 to 50 $\sunmass$. ii) The amount of Eu produced from a single event follows equation \ref{eqn:Random} with an average value $(M_0^{Eu})$ of $\num{2e-6}$ $\sunmass$. iii) The parameter $\alpha_{NS}$ depends on [Fe/H] and varying as equation \ref{alpha}; $\overline{\alpha}_{NS}$  is set to 0.275. iv) The coalescence time distribution of NSMs follows a DTD $\propto t^{-1.5}$. This systems has a minimum delay time of 1 Myr. The predictions of NSt3+$\alpha$ are plotted in left panel of Fig. \ref{fig:Model4}. It is seen that this model is in good agreement with the observational data, but it is not able to predict the presence of stars with low [Eu/Fe] (<$-$0.5) at [Fe/H] $< -$3.0. NSt3+$\alpha$ model predicts also the presence of stars with [Eu/Fe]< $-$0.5 even at relative high metallicity ([Fe/H]> $-$2.0)  that cannot be confirmed by the chosen data sample. Last, as seen in all the tested models of this work, the model cannot explain the presence of Eu in stars with [Fe/H]< $-$3.5 dex.

\begin{table*}
\centering
\renewcommand{\arraystretch}{1.2}
\begin{tabular}{c|ccc|ccc|ccc}
\multicolumn{1}{l|}{\multirow{3}{*}{{[}Fe/H{]} (dex)}} & \multicolumn{3}{c|}{Test1}          & \multicolumn{3}{c|}{Test2}          & \multicolumn{3}{c}{Test3}           \\
\multicolumn{1}{l|}{}                                  & mean {[}Eu/Fe{]} (dex) & sigma (dex) & $f$ & mean {[}Eu/Fe{]} (dex) & sigma (dex) & $f$ & mean {[}Eu/Fe{]} (dex) & sigma  & $f$ \\ \hline
$-$3.00                                                  & 1.42                   & 0.22     & 0.26  & 1.05                   & 0.23    & 0.17  & 0.84                   & 0.22     & 0.14  \\
$-$2.75                                                  & 1.01                   & 0.23     & 0.15  & 0.72                   & 0.23    & 0.04  & 0.59                   & 0.20     & 0.03  \\
$-$2.50                                                  & 0.69                   & 0.22     & 0.09  & 0.55                   & 0.19    & 0.02  & 0.50                   & 0.15     & 0.01  \\
$-$2.25                                                  & 0.53                   & 0.18     & 0.02  & 0.49                   & 0.14    & 0.00  & 0.47                   & 0.11     & 0.00 \\
$-$2.00                                                  & 0.47                   & 0.16     & 0.01  & 0.45                   & 0.11    & 0.00  & 0.45                   & 0.09     & 0.00 \\
$-$1.75                                                  & 0.43                   & 0.14     & 0.00  & 0.43                   & 0.10    & 0.00  & 0.42                   & 0.08     & 0.00 \\
$-$1.50                                                  & 0.37                   & 0.14     & 0.00  & 0.38                   & 0.10    & 0.00  & 0.37                   & 0.07     & 0.00 \\
$-$1.25                                                  & 0.28                   & 0.14     & 0.00  & 0.29                   & 0.10    & 0.00  & 0.30                   & 0.08     & 0.00 \\
$-$1.00                                                  & 0.15                   & 0.15     & 0.00  & 0.16                   & 0.10    & 0.00  & 0.17                   & 0.08     & 0.00
\end{tabular}
\caption{Here are summarise the results of our analysis. The table is organised as follows: in column 1, the metallicity value to which we compute mean and standard deviation of the [Eu/Fe] values, in column 2, mean [Eu/Fe] at some metallicity for a specific model, in column 3, the standard deviation at some metallicity for a specific model, in column 4, fraction of Eu-free ($f= N_{Eu-free} / N_{tot}$). This structure is repeated for the three different models contained in this section. }
\label{tab:dispertion}
\end{table*}
\noindent
Then, we tested a model with $\alpha_{NS}$ variable and the DTD $\propto t^{-1}$.
We built NSt1+$\alpha$ model with the following assumptions: i) Eu is produced only from NSM, whose progenitors are in the mass range from 9 to 50 $\sunmass$. ii) The amount of Eu produced from a single event follows equation \ref{eqn:Random} with an average value $(M_0^{Eu})$ of $\num{4e-6}$ $\sunmass$. iii) The parameter $\alpha_{NS}$ depends on [Fe/H] and varies as equation \ref{alpha}; $\overline{\alpha}_{NS}$  is set to 0.275. iv) The coalescence time of NSMs follows a DTD $\propto t^{-1}$ and has a minimum value of 1 Myr. The results of this model are shown in the right panel of Fig. \ref{fig:Model4}. As you can notice, the model does not predict the presence of stars with [Eu/Fe]< 0.3 at metallicity lower than $-$2.9 dex. On the other hand, the model shows good compatibility with the upper envelope of the stars' abundance distribution. We should also notice that in the region with [Eu/Fe] < $-$0.4 at metallicity larger than $-$2.5 dex, there is a strong excess in the predicted star distribution that is not supported by observational data. \\
By a simple comparison between the right panel of Fig. \ref{fig:Model4} and central panel of Fig. \ref{fig:Models1} we can assert that dropping the constancy of $\alpha_{NS}$ as a function of [Fe/H] has a great impact on the results of our chemical evolution models. In particular, an $\alpha_{NS}$ that varies with metallicity, can substitute MRD-SNe in the framework of explaining the low-Eu tail of metal-poor Halo stars ([Fe/H]<$-$2.5 dex).

\begin{figure}
  \includegraphics[width=\columnwidth, trim= 0cm 0cm 1cm 0cm, clip]{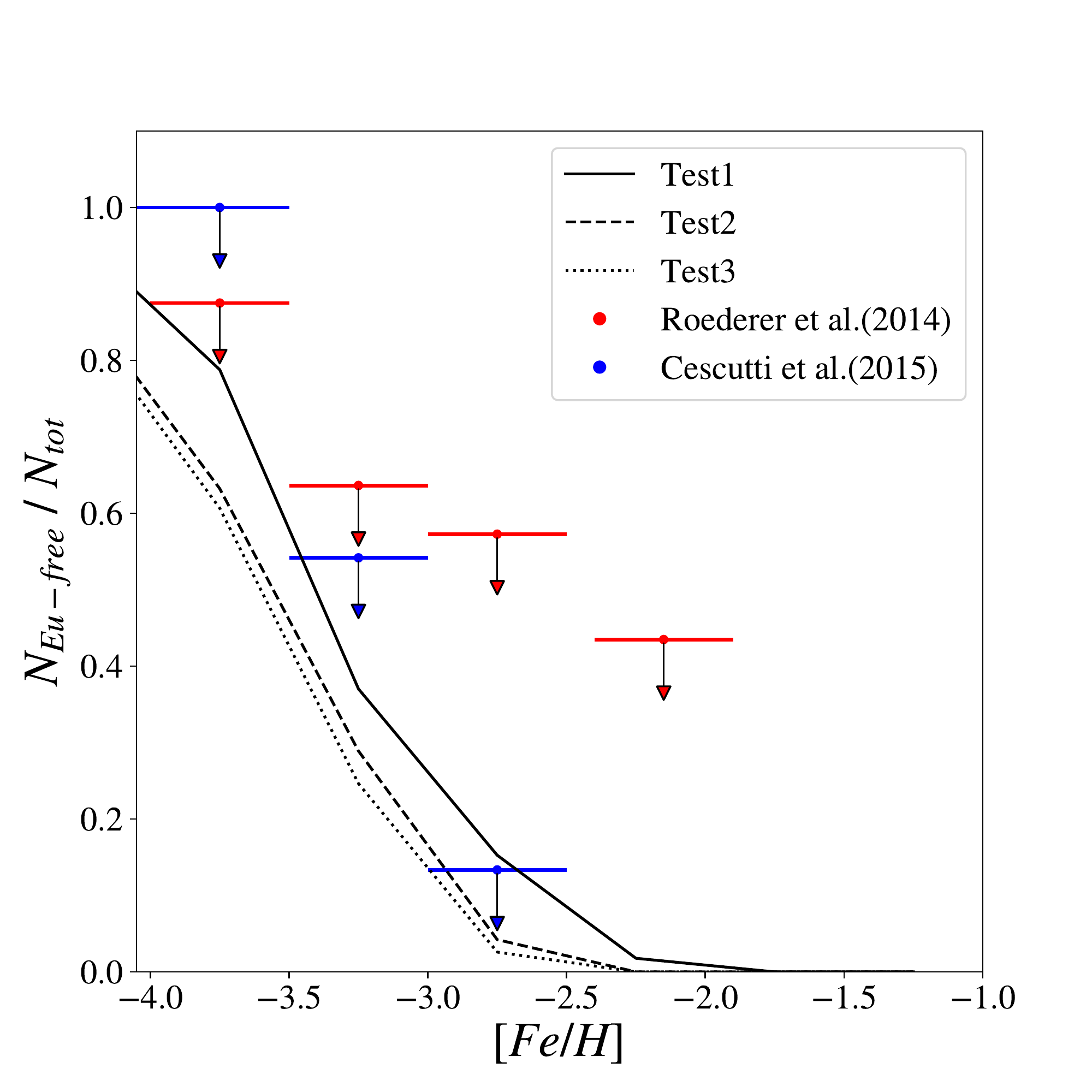}
  \caption{Ratio of Eu-free stars over the total number of stars for bins of 0.5 dex in [Fe/H]. In the figure are plotted the results for the  models Test1, Test2, and Test3. The blue triangles are the observational proxy for this ratio, so the ratio between the number of stars in which Eu only presents an upper limit (possibly Eu-free) over the number of stars for which at least Ba has been measured (total number of stars). The horizontal error bars show the dimension of each bin in [Fe/H]. Red triangles are the observational proxy for this ratio derived from the data-set used in this work, \citet{roederer2014search}. The blue triangles are the same observational proxy calculated in \citet{cescutti2015role}; adopting a different data-set \citep[cfr.][, for details on this data-set]{cescutti2015role}. }
  \label{fig:Eufree}
\end{figure}

\subsection{Test on the dispersion at intermediate metallicity}
In this section, we explore the correlation between the dispersion of the [Eu/Fe] values and the fraction of massive stars that can produce a NSM, namely the parameter $\alpha_{NS}$.\\
We start from the assumptions of our best model (NSt3+$\alpha$) and then we increase the value of $\alpha_{NS}$. With these prescriptions we create three different models (see Table \ref{tab:Models}): i) Test1 is exactly the same as NSt3+$\alpha$. ii) Test2; for this model, we assume
$\overline{\alpha}_{NS}$= 0.315 (see equation \ref{alpha}). This implies a slight increase of NSMs at the lowest metallicities, but it also implies an increase of a factor of 3 at [Fe/H]>$-$1.5. Due to this variation, the total number of NSM  is increased by a factor $\sim$2.5. For  this reason, since we want to keep approximately constant the total amount of Eu, we have to decrease $M_0^{Eu;NSM}$ to $\num{0.8e-6}$ $\sunmass$. iii) Test3; in this case we set $\overline{\alpha}_{NS}$= 0.375. As a consequence of this, the total number of NSM is increased by a factor $\sim$5.  Also, in this case, we reduce the mean Eu produced to $\num{0.4e-6}$ $\sunmass$. \\
On the results of these models, we select different bins in metallicity and in these we compute mean and standard deviation of [Eu/Fe] values. The mean and standard deviation for each model are reported in Table \ref{tab:dispertion}. In Fig. \ref{fig:Model5} are plotted the results of the three models: Test1, Test2, and Test3.\\
Looking at Fig. \ref{fig:Model5}, focusing at the region at the intermediate metallicity, it can be easily noticed that the observational data cannot exclude any of the tested models. Indeed, stochastic models with a large variation of $\alpha_{NS}$ (from 0.02 to 0.10) predict differently the enrichment at intermediate metallicity regime. On the other hand, the observational data are affected by relatively large uncertainties  (i.e. $\sim$ $0.2-0.3$ dex in [Eu/Fe]); moreover, the sample selected is certainly measured in a homogeneous way, but it is not large enough to apply safely a statistical approach. Adding more authors will increase the number of data, but we risk to increase significantly the scatter among different authors. Future surveys such as 4MOST \citep{4MOST} and WEAVE \citep{WEAVE} will surely produce larger data-set homogeneously measured and they could allow us to determine the value of $\alpha_{NS}$, and consequently  $M_0^{Eu;NSM}$, more precisely.

\subsubsection{Eu-free stars}
All the tested models have a common feature: the considered r-process events are rare and they are only a small fraction ($\alpha_{NS}$) of the total number of the main polluters of the ISM at low metallicity, the SNe II. It is easy to infer that, at extremely low metallicities ([Fe/H] $\leq -3$), a lot of low mass stars can be formed in regions where the ISM is not yet polluted by r-process events. We also expect that lowering the fraction  $\alpha_{NS}$ should lead to an increase of Eu-free stars (i.e. [Eu/Fe] $=-\infty$). Moreover, a longer time delay for the r-process events will also produce a higher fraction of Eu-free stars, since for a longer time ISM will be not enriched by r-process events. \\
All the plots of our models (Fig. \ref{fig:Models1}; \ref{fig:Models2}; \ref{fig:Models3}; \ref{fig:Model4}; \ref{fig:Model5}) show the long-living stars formed without Eu (i.e. Eu-free stars) at [Eu/Fe] $=-2.0$ dex. From these plots, it is possible already to find a behavior that is in agreement with our aforementioned expectations. 
For example, looking at Fig. \ref{fig:Models1}, is possible to notice that a model with a DTD function with the form $\propto t^{-1}$ (NSt1) predicts a higher number of Eu-free star compared to both the short and constant delay presented NS00 and the steeper DTD ($\propto t^{-1.5}$) of NSt3 models.
\\
However, to better study the behavior of the Eu-free stars with respect to the fraction  $\alpha_{NS}$, we report in Fig. \ref{fig:Eufree}, the ratio of Eu-free stars over the total number of stars for the models Test1, Test2, and Test3. In this plot, it appears clearly that increasing $\alpha_{NS}$, so moving from Test1 to Test3, the model predicts a lower fraction of Eu-free stars. 
Observationally, it is not obvious how to put constraints to the modeling since no Eu-free star has been yet claimed. On the other hand, several stars have only upper limits for europium. In Fig. \ref{fig:Eufree}, we decide to use as a proxy of Eu-free stars, stars for which only upper limits for europium have been detected and barium is measured, as already assumed in \citet{cescutti2015role}. 
In the plot, we use two data-set to compute this observational proxy. So together with the results obtained with the stars measured in \citet{roederer2014search}, we show also the results obtained in \citet{cescutti2015role} adopting a different data-set. Details of this collection can be found in \citet{cescutti2015role}. We decide to add these results, because it appears clear that the number of upper limits detected by \citet{roederer2014search} for europium are quite high and possibly due to a certain fraction of spectra missing the necessary quality to measure europium, rather than the real absence of this element. So, we trust more the results from the larger sample used in \citet{cescutti2015role} toward higher metallicity, whereas for low metallicity, they appear in reasonable agreement. In comparison, our models predictions appear to follow the trend, but it is always below the observational proxy. This can be explained by a  large fraction of false Eu-free stars, due to the difficulty of measuring the weak europium lines when the abundance is really low. Overall, our best model, (i.e. Test 1) appears in agreement with this observational proxy, but it is hard to find a firm conclusion from this prospective.

\section{Conclusions} \label{Conclusions}
In this paper, we have adopted the stochastic chemical evolution model of the Galactic halo presented by \citet{cescutti2008}, to study the impact of relaxing the constancy of the delay times for the coalescence of NSM, on the chemical evolution of Eu in the metal-poor environment of the Galactic halo. To perform that, we have implemented two different delay time distributions (DTDs) $\propto t^{-1}$ and $\propto t^{-1.5}$, as suggested in \citet{cote2019neutron}. For the Eu yields, we have followed the prescriptions of \citet{matteucci2014europium} and \citet{cescutti2015role}. We have also tried to find a way to solve the tensions between the observational data and the results of models that assume a variable coalescence time. In order to do that, we have explored a scenario in which both NSM and MRD SNe (magneto-rotationally drive SNe) are able to produce Eu. For the same reason, we have also implemented a fraction of massive stars that can produce  NSM systems that vary with metallicity, following the idea presented in \citet{simonetti2019new}. Finally, we have also studied the [Eu/Fe] vs. [Fe/H] in the Galactic halo and its correlation with the value of $\alpha_{NS}$.\\
Our main conclusions can be summarised as follows:\\\\


a) The NS-only scenario is in disagreement with observational data, even at moderate metallicity, when we assume a DTD$\propto t^{-1}$ for the coalescence timescales. On the other hand, assuming a DTD$\propto t^{-1.5}$ produces results similar to the ones with constant delay time. These conclusions are similar to the ones found by \citet{cote2019neutron}, but now we obtain these results in the framework of a stochastic chemical evolution model.  \\\\
b) The mixed scenario with NS and MRD SNe is able to explain the observed spread as shown, but only for a constant delay, in \citet{cescutti2015role}. The main assumptions, in this case, are that MRD SNe are 10 $\%$ of CC-SNe, explode only at low metallicity 
(Z<$10^{-3}$) and the production of Eu is the same for both NSM and MRD SNe.
We prove here that the models in this case
agree with observations independently by the assumed DTD.\\\\
c) Our best NS-only scenario is in good agreement with observational data under the following assumptions: i) Eu is produced only from NSM, whose progenitors are in the mass range from 9 to 50 $\sunmass$. ii) The amount of Eu produced from a single event follows equation \ref{eqn:Random} with an average value $(M_0^{Eu})$ of $\num{2e-6}$ $\sunmass$. iii) The parameter $\alpha_{NS}$ depends on [Fe/H] and varies as equation \ref{alpha}; the required $\overline{\alpha}_{NS}$  is 0.275. iv) The coalescence time distribution of NSMs should follow a DTD $\propto t^{-1.5}$ with a minimum value of 1 Myr. In this scenario, a larger fraction of NSM explodes in the early phases of the Galactic evolution, compared to nowadays \citep[see also][]{simonetti2019new}.\\\\
d) Adopting to our best model, we also show the predicted dispersion of [Eu/Fe] at a given metallicity depending on the \ref{alpha}; the comparison with the present literature data cannot allow us to put a stronger constraint \ref{alpha}. However, future high-resolution spectroscopical surveys, such as 4MOST \citep{4MOST} and WEAVE, \citep{WEAVE} will produce the necessary statistic to constrain at best this parameter.\\\\
e) Our best model is in agreement with the chosen observational proxy for Eu-free stars. However, the fraction of false Eu free stars cannot be evaluated and no firm conclusions can be raised.\\\\
\noindent
The models struggle to reproduce the low-metallicity tail of stars with [Eu/Fe]<$-0.1$ dex at [Fe/H]<$-3.0$ dex. We underline that the model at this stage does not consider several complexities that can play an important role to solve this issue, for example: stochasticity in SFR and infall-law, cross-contamination of sub-haloes, pre-enrichment of the infalling gas, and multi-phase ISM.\\
In a future work, we will try to solve this problem by taking into account the hierarchical formation of Galactic halo by accretion of satellite galaxies. In fact, the enrichment of r-process elements in these objects could have been less effective due to dynamical effects 
connected to the formation of binary neutron stars \citep[see][]{bonetti2019neutron}.

\section*{Acknowledgements}
This work has been partially supported by the European Union Cooperation in Science and Technology (COST) Action CA16117 (ChETEC). We thank Paolo Simonetti for assistance with the implementation of the time delay distributions.

\clearpage




\bibliographystyle{mnras}
\bibliography{example} 

\begin{thebibliography}{}
\makeatletter
\relax
\def\mn@urlcharsother{\let\do\@makeother \do\$\do\&\do\#\do\^\do\_\do\%\do\~}
\def\mn@doi{\begingroup\mn@urlcharsother \@ifnextchar [ {\mn@doi@}
  {\mn@doi@[]}}
\def\mn@doi@[#1]#2{\def\@tempa{#1}\ifx\@tempa\@empty \href
  {http://dx.doi.org/#2} {doi:#2}\else \href {http://dx.doi.org/#2} {#1}\fi
  \endgroup}
\def\mn@eprint#1#2{\mn@eprint@#1:#2::\@nil}
\def\mn@eprint@arXiv#1{\href {http://arxiv.org/abs/#1} {{\tt arXiv:#1}}}
\def\mn@eprint@dblp#1{\href {http://dblp.uni-trier.de/rec/bibtex/#1.xml}
  {dblp:#1}}
\def\mn@eprint@#1:#2:#3:#4\@nil{\def\@tempa {#1}\def\@tempb {#2}\def\@tempc
  {#3}\ifx \@tempc \@empty \let \@tempc \@tempb \let \@tempb \@tempa \fi \ifx
  \@tempb \@empty \def\@tempb {arXiv}\fi \@ifundefined
  {mn@eprint@\@tempb}{\@tempb:\@tempc}{\expandafter \expandafter \csname
  mn@eprint@\@tempb\endcsname \expandafter{\@tempc}}}

\bibitem[\protect\citeauthoryear{Abbott et~al.,}{Abbott
  et~al.}{2017a}]{abbott2017search}
Abbott B.~P.,  et~al., 2017a, Physical Review D, 95, 082005

\bibitem[\protect\citeauthoryear{Abbott et~al.,}{Abbott
  et~al.}{2017b}]{abbott2017gw170817}
Abbott B.~P.,  et~al., 2017b, Physical Review Letters, 119, 161101

\bibitem[\protect\citeauthoryear{Abohalima \& Frebel}{Abohalima \&
  Frebel}{2018}]{Abohalima_2018}
Abohalima A.,  Frebel A.,  2018, \mn@doi [The Astrophysical Journal Supplement
  Series] {10.3847/1538-4365/aadfe9}, 238, 36

\bibitem[\protect\citeauthoryear{Arcones \& Thielemann}{Arcones \&
  Thielemann}{2012}]{Arcones_2012}
Arcones A.,  Thielemann F.-K.,  2012, \mn@doi [Journal of Physics G: Nuclear
  and Particle Physics] {10.1088/0954-3899/40/1/013201}, 40, 013201

\bibitem[\protect\citeauthoryear{Arcones, Janka  \& Scheck}{Arcones
  et~al.}{2007}]{arcones2007nucleosynthesis}
Arcones A.,  Janka H.-T.,   Scheck L.,  2007, Astronomy \& Astrophysics, 467,
  1227

\bibitem[\protect\citeauthoryear{Argast, Samland, Thielemann  \& Qian}{Argast
  et~al.}{2004}]{argast2004neutron}
Argast D.,  Samland M.,  Thielemann F.-K.,   Qian Y.-Z.,  2004, Astronomy \&
  Astrophysics, 416, 997

\bibitem[\protect\citeauthoryear{{Bauswein}, {Goriely}  \& {Janka}}{{Bauswein}
  et~al.}{2013}]{2013ApJ...773...78B}
{Bauswein} A.,  {Goriely} S.,   {Janka} H.~T.,  2013, \mn@doi [\apj]
  {10.1088/0004-637X/773/1/78}, \href
  {https://ui.adsabs.harvard.edu/abs/2013ApJ...773...78B} {773, 78}

\bibitem[\protect\citeauthoryear{Berger}{Berger}{2014}]{berger2014short}
Berger E.,  2014, Annual review of Astronomy and Astrophysics, 52, 43

\bibitem[\protect\citeauthoryear{{Bisterzo} et~al.,}{{Bisterzo}
  et~al.}{2015}]{2015MNRAS.449..506B}
{Bisterzo} S.,  et~al., 2015, \mn@doi [\mnras] {10.1093/mnras/stv271}, \href
  {https://ui.adsabs.harvard.edu/abs/2015MNRAS.449..506B} {449, 506}

\bibitem[\protect\citeauthoryear{Bogomazov, Lipunov  \& Tutukov}{Bogomazov
  et~al.}{2007}]{bogomazov2007evolution}
Bogomazov A.,  Lipunov V.,   Tutukov A.,  2007, Astronomy Reports, 51, 308

\bibitem[\protect\citeauthoryear{{Bonetti}, {Perego}, {Dotti}  \&
  {Cescutti}}{{Bonetti} et~al.}{2019}]{bonetti2019neutron}
{Bonetti} M.,  {Perego} A.,  {Dotti} M.,   {Cescutti} G.,  2019, \mn@doi
  [\mnras] {10.1093/mnras/stz2554}, \href
  {https://ui.adsabs.harvard.edu/abs/2019MNRAS.490..296B} {490, 296}

\bibitem[\protect\citeauthoryear{Cameron}{Cameron}{1982}]{cameron1982elemental}
Cameron A.,  1982, Essays in Nuclear Astrophysics, pp 23--43

\bibitem[\protect\citeauthoryear{Cescutti}{Cescutti}{2008}]{cescutti2008}
Cescutti G.,  2008, Astronomy \& Astrophysics, 481, 691

\bibitem[\protect\citeauthoryear{Cescutti \& Chiappini}{Cescutti \&
  Chiappini}{2014}]{cescutti2014explaining}
Cescutti G.,  Chiappini C.,  2014, Astronomy \& Astrophysics, 565, A51

\bibitem[\protect\citeauthoryear{Cescutti, Fran{\c{c}}ois, Matteucci, Cayrel
  \& Spite}{Cescutti et~al.}{2006}]{cescutti2006chemical}
Cescutti G.,  Fran{\c{c}}ois P.,  Matteucci F.,  Cayrel R.,   Spite M.,  2006,
  Astronomy \& Astrophysics, 448, 557

\bibitem[\protect\citeauthoryear{Cescutti, Romano, Matteucci, Chiappini  \&
  Hirschi}{Cescutti et~al.}{2015}]{cescutti2015role}
Cescutti G.,  Romano D.,  Matteucci F.,  Chiappini C.,   Hirschi R.,  2015,
  Astronomy \& Astrophysics, 577, A139

\bibitem[\protect\citeauthoryear{Chiappini, Ekstr{\"o}m, Meynet, Hirschi,
  Maeder  \& Charbonnel}{Chiappini et~al.}{2008}]{chiappini2008new}
Chiappini C.,  Ekstr{\"o}m S.,  Meynet G.,  Hirschi R.,  Maeder A.,
  Charbonnel C.,  2008, Astronomy \& Astrophysics, 479, L9

\bibitem[\protect\citeauthoryear{Chruslinska, Belczynski, Klencki  \&
  Benacquista}{Chruslinska et~al.}{2018}]{chruslinska2018double}
Chruslinska M.,  Belczynski K.,  Klencki J.,   Benacquista M.,  2018, Monthly
  Notices of the Royal Astronomical Society, 474, 2937

\bibitem[\protect\citeauthoryear{C{\^o}t{\'e} et~al.,}{C{\^o}t{\'e}
  et~al.}{2019}]{cote2019neutron}
C{\^o}t{\'e} B.,  et~al., 2019, The Astrophysical Journal, 875, 106

\bibitem[\protect\citeauthoryear{Coulter et~al.,}{Coulter
  et~al.}{2017}]{coulter2017swope}
Coulter D.,  et~al., 2017, Science, 358, 1556

\bibitem[\protect\citeauthoryear{Cowan, Thielemann  \& Truran}{Cowan
  et~al.}{1991}]{cowan1991r}
Cowan J.~J.,  Thielemann F.-K.,   Truran J.~W.,  1991, Physics Reports, 208,
  267

\bibitem[\protect\citeauthoryear{D'Avanzo}{D'Avanzo}{2015}]{d2015short}
D'Avanzo P.,  2015, Journal of High Energy Astrophysics, 7, 73

\bibitem[\protect\citeauthoryear{{Dalton} et~al.,}{{Dalton}
  et~al.}{2012}]{WEAVE}
{Dalton} G.,  et~al., 2012, in Ground-based and Airborne Instrumentation for
  Astronomy IV. p. 84460P, \mn@doi{10.1117/12.925950}

\bibitem[\protect\citeauthoryear{Dominik, Belczynski, Fryer, Holz, Berti,
  Bulik, Mandel  \& O{\textquotesingle}Shaughnessy}{Dominik
  et~al.}{2012}]{Dominik_2012}
Dominik M.,  Belczynski K.,  Fryer C.,  Holz D.~E.,  Berti E.,  Bulik T.,
  Mandel I.,   O{\textquotesingle}Shaughnessy R.,  2012, \mn@doi [The
  Astrophysical Journal] {10.1088/0004-637x/759/1/52}, 759, 52

\bibitem[\protect\citeauthoryear{{Eichler}, {Livio}, {Piran}  \&
  {Schramm}}{{Eichler} et~al.}{1989}]{1989Natur.340..126E}
{Eichler} D.,  {Livio} M.,  {Piran} T.,   {Schramm} D.~N.,  1989, \mn@doi
  [\nat] {10.1038/340126a0}, \href
  {https://ui.adsabs.harvard.edu/abs/1989Natur.340..126E} {340, 126}

\bibitem[\protect\citeauthoryear{Fong et~al.,}{Fong et~al.}{2017}]{Fong_2017}
Fong W.,  et~al., 2017, \mn@doi [The Astrophysical Journal]
  {10.3847/2041-8213/aa9018}, 848, L23

\bibitem[\protect\citeauthoryear{Freiburghaus, Rosswog  \&
  Thielemann}{Freiburghaus et~al.}{1999}]{freiburghaus1999r}
Freiburghaus C.,  Rosswog S.,   Thielemann F.-K.,  1999, The Astrophysical
  Journal Letters, 525, L121

\bibitem[\protect\citeauthoryear{Fulbright}{Fulbright}{2000}]{fulbright2000abundances}
Fulbright J.~P.,  2000, The Astronomical Journal, 120, 1841

\bibitem[\protect\citeauthoryear{Giacobbo \& Mapelli}{Giacobbo \&
  Mapelli}{2018}]{giacobbo2018progenitors}
Giacobbo N.,  Mapelli M.,  2018, Monthly Notices of the Royal Astronomical
  Society, 480, 2011

\bibitem[\protect\citeauthoryear{Graur et~al.,}{Graur
  et~al.}{2011}]{graur2011supernovae}
Graur O.,  et~al., 2011, Monthly Notices of the Royal Astronomical Society,
  417, 916

\bibitem[\protect\citeauthoryear{Heringer, Pritchet, Kezwer, Graham, Sand  \&
  Bildfell}{Heringer et~al.}{2016}]{heringer2016type}
Heringer E.,  Pritchet C.,  Kezwer J.,  Graham M.~L.,  Sand D.,   Bildfell C.,
  2016, The Astrophysical Journal, 834, 15

\bibitem[\protect\citeauthoryear{Honda, Aoki, Kajino, Ando, Beers, Izumiura,
  Sadakane  \& Takada-Hidai}{Honda et~al.}{2004}]{honda2004spectroscopic}
Honda S.,  Aoki W.,  Kajino T.,  Ando H.,  Beers T.~C.,  Izumiura H.,  Sadakane
  K.,   Takada-Hidai M.,  2004, The Astrophysical Journal, 607, 474

\bibitem[\protect\citeauthoryear{{Hotokezaka}, {Kiuchi}, {Kyutoku},
  {Muranushi}, {Sekiguchi}, {Shibata}  \& {Taniguchi}}{{Hotokezaka}
  et~al.}{2013}]{2013PhRvD..88d4026H}
{Hotokezaka} K.,  {Kiuchi} K.,  {Kyutoku} K.,  {Muranushi} T.,  {Sekiguchi}
  Y.-i.,  {Shibata} M.,   {Taniguchi} K.,  2013, \mn@doi [\prd]
  {10.1103/PhysRevD.88.044026}, \href
  {https://ui.adsabs.harvard.edu/abs/2013PhRvD..88d4026H} {88, 044026}

\bibitem[\protect\citeauthoryear{Howard, Mathews, Takahashi  \& Ward}{Howard
  et~al.}{1986}]{howard1986parametric}
Howard W.,  Mathews G.,  Takahashi K.,   Ward R.,  1986, The Astrophysical
  Journal, 309, 633

\bibitem[\protect\citeauthoryear{Ishimaru, Wanajo, Aoki  \& Ryan}{Ishimaru
  et~al.}{2003}]{ishimaru2003detection}
Ishimaru Y.,  Wanajo S.,  Aoki W.,   Ryan S.~G.,  2003, The Astrophysical
  Journal Letters, 600, L47

\bibitem[\protect\citeauthoryear{Ivanova, Heinke, Rasio, Belczynski  \&
  Fregeau}{Ivanova et~al.}{2008}]{ivanova2008formation}
Ivanova N.,  Heinke C.,  Rasio F.~A.,  Belczynski K.,   Fregeau J.,  2008,
  Monthly Notices of the Royal Astronomical Society, 386, 553

\bibitem[\protect\citeauthoryear{Kalogera et~al.,}{Kalogera
  et~al.}{2004}]{kalogera2004cosmic}
Kalogera V.,  et~al., 2004, The Astrophysical Journal Letters, 601, L179

\bibitem[\protect\citeauthoryear{{Karlsson, T.} \& {Gustafsson, B.}}{{Karlsson,
  T.} \& {Gustafsson, B.}}{2005}]{refId0}
{Karlsson, T.} {Gustafsson, B.} 2005, \mn@doi [A\&A]
  {10.1051/0004-6361:20042168}, 436, 879

\bibitem[\protect\citeauthoryear{Koch \& Edvardsson}{Koch \&
  Edvardsson}{2002}]{koch2002europium}
Koch A.,  Edvardsson B.,  2002, Astronomy \& Astrophysics, 381, 500

\bibitem[\protect\citeauthoryear{{Komiya} \& {Shigeyama}}{{Komiya} \&
  {Shigeyama}}{2016}]{2016ApJ...830...76K}
{Komiya} Y.,  {Shigeyama} T.,  2016, \mn@doi [\apj]
  {10.3847/0004-637X/830/2/76}, \href
  {https://ui.adsabs.harvard.edu/abs/2016ApJ...830...76K} {830, 76}

\bibitem[\protect\citeauthoryear{Korobkin, Rosswog, Arcones  \&
  Winteler}{Korobkin et~al.}{2012}]{korobkin2012astrophysical}
Korobkin O.,  Rosswog S.,  Arcones A.,   Winteler C.,  2012, Monthly Notices of
  the Royal Astronomical Society, 426, 1940

\bibitem[\protect\citeauthoryear{Maeder \& Meynet}{Maeder \&
  Meynet}{1989}]{maeder1989grids}
Maeder A.,  Meynet G.,  1989, Astronomy and Astrophysics, 210, 155

\bibitem[\protect\citeauthoryear{Maoz \& Badenes}{Maoz \&
  Badenes}{2010}]{maoz2010supernova}
Maoz D.,  Badenes C.,  2010, Monthly Notices of the Royal Astronomical Society,
  407, 1314

\bibitem[\protect\citeauthoryear{Maoz \& Mannucci}{Maoz \&
  Mannucci}{2012}]{maoz2012type}
Maoz D.,  Mannucci F.,  2012, Publications of the Astronomical Society of
  Australia, 29, 447

\bibitem[\protect\citeauthoryear{{Matteucci} \& {Greggio}}{{Matteucci} \&
  {Greggio}}{1986}]{1986A&A...154..279M}
{Matteucci} F.,  {Greggio} L.,  1986, \aap, \href
  {https://ui.adsabs.harvard.edu/abs/1986A&A...154..279M} {154, 279}

\bibitem[\protect\citeauthoryear{Matteucci, Panagia, Pipino, Mannucci, Recchi
  \& Della~Valle}{Matteucci et~al.}{2006}]{10.1111/j.1365-2966.2006.10848.x}
Matteucci F.,  Panagia N.,  Pipino A.,  Mannucci F.,  Recchi S.,   Della~Valle
  M.,  2006, \mn@doi [Monthly Notices of the Royal Astronomical Society]
  {10.1111/j.1365-2966.2006.10848.x}, 372, 265

\bibitem[\protect\citeauthoryear{Matteucci, Romano, Arcones, Korobkin  \&
  Rosswog}{Matteucci et~al.}{2014}]{matteucci2014europium}
Matteucci F.,  Romano D.,  Arcones A.,  Korobkin O.,   Rosswog S.,  2014,
  Monthly Notices of the Royal Astronomical Society, 438, 2177

\bibitem[\protect\citeauthoryear{{McWilliam}}{{McWilliam}}{1998}]{mcwilliam98}
{McWilliam} A.,  1998, \mn@doi [\aj] {10.1086/300289}, \href
  {https://ui.adsabs.harvard.edu/abs/1998AJ....115.1640M} {115, 1640}

\bibitem[\protect\citeauthoryear{Mennekens \& Vanbeveren}{Mennekens \&
  Vanbeveren}{2014}]{mennekens2014massive}
Mennekens N.,  Vanbeveren D.,  2014, Astronomy \& Astrophysics, 564, A134

\bibitem[\protect\citeauthoryear{{M{\"o}sta}, {Ott}, {Radice}, {Roberts},
  {Schnetter}  \& {Haas}}{{M{\"o}sta} et~al.}{2015}]{2015Natur.528..376M}
{M{\"o}sta} P.,  {Ott} C.~D.,  {Radice} D.,  {Roberts} L.~F.,  {Schnetter} E.,
   {Haas} R.,  2015, \mn@doi [\nat] {10.1038/nature15755}, \href
  {https://ui.adsabs.harvard.edu/abs/2015Natur.528..376M} {528, 376}

\bibitem[\protect\citeauthoryear{{Nishimura}, {Takiwaki}  \&
  {Thielemann}}{{Nishimura} et~al.}{2015}]{2015ApJ...810..109N}
{Nishimura} N.,  {Takiwaki} T.,   {Thielemann} F.-K.,  2015, \mn@doi [\apj]
  {10.1088/0004-637X/810/2/109}, \href
  {https://ui.adsabs.harvard.edu/abs/2015ApJ...810..109N} {810, 109}

\bibitem[\protect\citeauthoryear{{Oechslin}, {Janka}  \& {Marek}}{{Oechslin}
  et~al.}{2007}]{2007A&A...467..395O}
{Oechslin} R.,  {Janka} H.~T.,   {Marek} A.,  2007, \mn@doi [\aap]
  {10.1051/0004-6361:20066682}, \href
  {https://ui.adsabs.harvard.edu/abs/2007A&A...467..395O} {467, 395}

\bibitem[\protect\citeauthoryear{{Panov}, {Korneev}  \& {Thielemann}}{{Panov}
  et~al.}{2008}]{2008AstL...34..189P}
{Panov} I.~V.,  {Korneev} I.~Y.,   {Thielemann} F.~K.,  2008, \mn@doi
  [Astronomy Letters] {10.1007/s11443-008-3006-1}, \href
  {https://ui.adsabs.harvard.edu/abs/2008AstL...34..189P} {34, 189}

\bibitem[\protect\citeauthoryear{{Perego}, {Rosswog}, {Cabez{\'o}n},
  {Korobkin}, {K{\"a}ppeli}, {Arcones}  \& {Liebend{\"o}rfer}}{{Perego}
  et~al.}{2014}]{2014MNRAS.443.3134P}
{Perego} A.,  {Rosswog} S.,  {Cabez{\'o}n} R.~M.,  {Korobkin} O.,
  {K{\"a}ppeli} R.,  {Arcones} A.,   {Liebend{\"o}rfer} M.,  2014, \mn@doi
  [\mnras] {10.1093/mnras/stu1352}, \href
  {https://ui.adsabs.harvard.edu/abs/2014MNRAS.443.3134P} {443, 3134}

\bibitem[\protect\citeauthoryear{Rodney et~al.,}{Rodney
  et~al.}{2014}]{rodney2014type}
Rodney S.~A.,  et~al., 2014, The Astronomical Journal, 148, 13

\bibitem[\protect\citeauthoryear{Roederer, Preston, Thompson, Shectman, Sneden,
  Burley  \& Kelson}{Roederer et~al.}{2014}]{roederer2014search}
Roederer I.~U.,  Preston G.~W.,  Thompson I.~B.,  Shectman S.~A.,  Sneden C.,
  Burley G.~S.,   Kelson D.~D.,  2014, The Astronomical Journal, 147, 136

\bibitem[\protect\citeauthoryear{Scalo}{Scalo}{1986}]{scalo1986stellar}
Scalo J.~M.,  1986, Fundamentals of cosmic physics, 11, 1

\bibitem[\protect\citeauthoryear{Sch{\"o}nrich \& Weinberg}{Sch{\"o}nrich \&
  Weinberg}{2019}]{schonrich2019chemical}
Sch{\"o}nrich R.~A.,  Weinberg D.~H.,  2019, Monthly Notices of the Royal
  Astronomical Society, 487, 580

\bibitem[\protect\citeauthoryear{Shao \& Li}{Shao \& Li}{2018}]{shao2018black}
Shao Y.,  Li X.-D.,  2018, Monthly Notices of the Royal Astronomical Society:
  Letters, 477, L128

\bibitem[\protect\citeauthoryear{Shen, Cooke, Ramirez-Ruiz, Madau, Mayer  \&
  Guedes}{Shen et~al.}{2015}]{Shen_2015}
Shen S.,  Cooke R.~J.,  Ramirez-Ruiz E.,  Madau P.,  Mayer L.,   Guedes J.,
  2015, \mn@doi [The Astrophysical Journal] {10.1088/0004-637x/807/2/115}, 807,
  115

\bibitem[\protect\citeauthoryear{Siegel, Barnes  \& Metzger}{Siegel
  et~al.}{2019}]{Siegel_2019}
Siegel D.~M.,  Barnes J.,   Metzger B.~D.,  2019, \mn@doi [Nature]
  {10.1038/s41586-019-1136-0}, 569, 241–244

\bibitem[\protect\citeauthoryear{Simonetti, Matteucci, Greggio  \&
  Cescutti}{Simonetti et~al.}{2019}]{simonetti2019new}
Simonetti P.,  Matteucci F.,  Greggio L.,   Cescutti G.,  2019, Monthly Notices
  of the Royal Astronomical Society, 486, 2896

\bibitem[\protect\citeauthoryear{{Symbalisty} \& {Schramm}}{{Symbalisty} \&
  {Schramm}}{1982}]{1982ApL....22..143S}
{Symbalisty} E.,  {Schramm} D.~N.,  1982, \aplett, \href
  {https://ui.adsabs.harvard.edu/abs/1982ApL....22..143S} {22, 143}

\bibitem[\protect\citeauthoryear{{Tanvir}, {Levan}, {Fruchter}, {Hjorth},
  {Hounsell}, {Wiersema}  \& {Tunnicliffe}}{{Tanvir}
  et~al.}{2013}]{2013Natur.500..547T}
{Tanvir} N.~R.,  {Levan} A.~J.,  {Fruchter} A.~S.,  {Hjorth} J.,  {Hounsell}
  R.~A.,  {Wiersema} K.,   {Tunnicliffe} R.~L.,  2013, \mn@doi [\nat]
  {10.1038/nature12505}, \href
  {https://ui.adsabs.harvard.edu/abs/2013Natur.500..547T} {500, 547}

\bibitem[\protect\citeauthoryear{Tauris et~al.,}{Tauris
  et~al.}{2017}]{tauris2017formation}
Tauris T.,  et~al., 2017, The Astrophysical Journal, 846, 170

\bibitem[\protect\citeauthoryear{Thornton, Gaudlitz, Janka  \&
  Steinmetz}{Thornton et~al.}{1998}]{thornton1998energy}
Thornton K.,  Gaudlitz M.,  Janka H.-T.,   Steinmetz M.,  1998, The
  Astrophysical Journal, 500, 95

\bibitem[\protect\citeauthoryear{Totani, Morokuma, Oda, Doi  \& Yasuda}{Totani
  et~al.}{2008}]{totani2008delay}
Totani T.,  Morokuma T.,  Oda T.,  Doi M.,   Yasuda N.,  2008, Publications of
  the Astronomical Society of Japan, 60, 1327

\bibitem[\protect\citeauthoryear{Wanajo, Kajino, Mathews  \& Otsuki}{Wanajo
  et~al.}{2001}]{wanajo2001r}
Wanajo S.,  Kajino T.,  Mathews G.~J.,   Otsuki K.,  2001, The Astrophysical
  Journal, 554, 578

\bibitem[\protect\citeauthoryear{{Wanajo}, {Janka}, {M{\"u}ller}  \&
  {Kubono}}{{Wanajo} et~al.}{2011}]{2011JPhCS.312d2008W}
{Wanajo} S.,  {Janka} H.~T.,  {M{\"u}ller} B.,   {Kubono} S.,  2011, in Journal
  of Physics Conference Series. p. 042008,
  \mn@doi{10.1088/1742-6596/312/4/042008}

\bibitem[\protect\citeauthoryear{Wanajo, Sekiguchi, Nishimura, Kiuchi, Kyutoku
  \& Shibata}{Wanajo et~al.}{2014}]{Wanajo_2014}
Wanajo S.,  Sekiguchi Y.,  Nishimura N.,  Kiuchi K.,  Kyutoku K.,   Shibata M.,
   2014, \mn@doi [The Astrophysical Journal] {10.1088/2041-8205/789/2/l39},
  789, L39

\bibitem[\protect\citeauthoryear{Winteler, Kaeppeli, Perego, Arcones, Vasset,
  Nishimura, Liebendoerfer  \& Thielemann}{Winteler
  et~al.}{2012}]{winteler2012magnetorotationally}
Winteler C.,  Kaeppeli R.,  Perego A.,  Arcones A.,  Vasset N.,  Nishimura N.,
  Liebendoerfer M.,   Thielemann F.-K.,  2012, The astrophysical journal
  letters, 750, L22

\bibitem[\protect\citeauthoryear{Woolf, Tomkin  \& Lambert}{Woolf
  et~al.}{1995}]{woolf1995r}
Woolf V.~M.,  Tomkin J.,   Lambert D.~L.,  1995, The Astrophysical Journal,
  453, 660

\bibitem[\protect\citeauthoryear{Woosley, Wilson, Mathews, Hoffman  \&
  Meyer}{Woosley et~al.}{1994}]{woosley1994r}
Woosley S.,  Wilson J.,  Mathews G.,  Hoffman R.,   Meyer B.,  1994, The
  Astrophysical Journal, 433, 229

\bibitem[\protect\citeauthoryear{{Yoon}, {Langer}  \& {Norman}}{{Yoon}
  et~al.}{2006}]{2006A&A...460..199Y}
{Yoon} S.~C.,  {Langer} N.,   {Norman} C.,  2006, \mn@doi [\aap]
  {10.1051/0004-6361:20065912}, \href
  {https://ui.adsabs.harvard.edu/abs/2006A&A...460..199Y} {460, 199}

\bibitem[\protect\citeauthoryear{{de Jong} et~al.,}{{de Jong}
  et~al.}{2014}]{4MOST}
{de Jong} R.~S.,  et~al., 2014, in Ground-based and Airborne Instrumentation
  for Astronomy V. p. 91470M, \mn@doi{10.1117/12.2055826}

\makeatother
\end{thebibliography}

\bsp	
\label{lastpage}
\end{document}